# What enables GaOx as hole transport layer for a 16% 1.0 eV $CuInSe_2$ Bottom Cells with $V_{OC}$ above 550 mV?

*Francesco Lodola*, Zhuangyi Zhou, Boaz Koren, Saeed Bayat, Alessandro Magon, Yucheng Hu, Adrian-Marie Philippe, Michele Melchiorre, Hasan Arif Yetkin, Gunnar Kusch, Nathalie Valle, Rachel A. Oliver and Susanne Siebentritt**

Francesco Lodola, Zhuangyi Zhou, Boaz Koren, Saeed Bayat, Alessandro Magon, Michele Melchiorre, Hasan Arif Yetkin, Susanne Siebentritt
University of Luxembourg, Department of Physics and Materials Science, 41 Rue du Brill, Belvaux, L-4422 Luxembourg

Yucheng Hu, Gunnar Kusch, Rachel A. Oliver
Department of Materials Science and Metallurgy, University of Cambridge, 27 Charles Babbage Road, Cambridge CB3 0FS, United Kingdom

Adrian-Marie Philippe, Nathalie Valle
Luxembourg Institute of Science and Technology (LIST), 41, rue du Brill, L-4422, Belvaux, Luxembourg

*E-mail: francesco.lodola@uni.lu, susanne.siebentritt@uni.lu

**Abstract**

Among the highly efficient photovoltaic technologies, that do not rely on epitaxy, only chalcopyrites have a bandgap tunable down to 1.00 eV, the ideal for tandem applications. This is obtained with a pure $CuInSe_2$ absorber without Ga. $GaO_x$ has been shown to be an efficient hole transport layer that prevents recombination at the metallic back contact. On the other hand, $GaO_x$ has proven detrimental, when it forms on In containing transparent back contacts in bifacial solar cells. Here, we investigate the conditions that make the $GaO_x$ layer conductive. We employ a GaOx hole transport layer that is formed through ion exchange during co-evaporation of the low band gap absorber layer. We find that no additional Cu is needed, and that Na is not necessary for a conductive $GaO_x$. Nor did we find a systematic influence of oxygen flow during the sputtering process of the oxide layer. The $GaO_x$ layer is partly crystalline. The optimized passivating hole transport layer enables a $CuInSe_2$ bottom solar cell, without any addition of Ag or heavy alkalis, with an active area efficiency above 16% and a record-certified open-circuit voltage $V_{OC}$ of 552meV.

## 1. Introduction

Photovoltaics (PV) play a key role in reducing greenhouse gas emissions, while providing clean electricity for the long-term development of a sustainable society.[1] Recently, tandem PV technologies have drawn a lot of attention, because they enable the possibility of exceeding the Shockley-Queisser limit for single-junction solar cells.[2] In fact, tandem cells can achieve efficiencies above 46% due to improved solar spectrum utilisation.[3] For the top cell, wide bandgap perovskites have been the most investigated candidates, while different options for the bottom cell have been employed: chalcopyrite[4–7], Si[8,9], perovskite[10] and organic cells[11]. Among all these technologies, chalcopyrite solar cells stand out for their tuneable bandgap that can be adjusted down to 1.00 eV,[12] which allows for optimal utilization of the solar spectrum, leading to higher efficiencies. In addition, chalcopyrite solar cells have achieved high efficiencies exceeding 23%,[13,14] for the standard bandgap ($E_g \geq 1.08$ eV), and 20%,[15,16] for the low bandgap ($E_g \leq 1.03$ eV). Moreover, $Cu(InGa)Se_2$ solar cells have proven to be radiation-resistant, making them particularly attractive for space applications.[17,18] They can also be manufactured on flexible and light weight substrates, even in monolithic tandem configurations maintaining efficiencies above 23%[19,20] after bending cycles, opening up possibilities for fully flexible devices.

In order to achieve the narrowest bandgap of 1.00 eV, the chalcopyrite absorber needs to be pure $CuInSe_2$, without Ga. However, this results in solar cells suffering from low open-circuit voltage ($V_{OC}$), due to high recombination velocities at the molybdenum back contact.[21] Indeed, such devices suffer from high non-radiative losses and the best power conversion efficiencies (PCE) achieved so far are 16.0% with a $V_{OC}$ of 526 mV,[22] and 15.0% with a $V_{OC}$ of 491 mV,[23] with and without potassium fluoride treatment respectively. To avoid incorporating a Ga gradient towards the back contact, which mitigates non-radiative losses[24,25] but introduces generation and radiative losses,[26–28] a hole transport layer (HTL) has recently been developed for chalcopyrite solar cells.[29–31] This HTL was grown as a 100 nm co-evaporated $CuGaSe_2$ layer, followed by 40 nm $InO_x$. After the absorber growth, the two layers of the HTL show complete ion-exchange between Ga and In, resulting in $GaO_x$ on top of a $CuInSe_2$ thin layer. Originally, the InOx was solution processed. These HTLs with remarkable transport properties, leading to fill factors (*FF*) above 70%, were obtained when a Cu precursor (around 30-40 nm) was supplied and annealed at 500 °C for 20 minutes, before the standard 3-stage co-evaporation $CuInSe_2$ absorber growth. The Cu precursor was thought to facilitate the formation of deep Cu-related defects in the oxide layer, forming a band that allows holes to be collected at the Mo back contact.[32–36] In fact, this HTL, in which the oxide layer was deposited by a solution method, has shown a good passivation effect but a blocking behaviour, resulting in a poor *FF*, if it was not annealed with additional Cu prior to absorber growth.[32,36]

The solution-processed $InO_x$ layer on top the of $CuGaSe_2$ has demonstrated thermal stability under the absorber growth condition in the sense that it did not diffuse or dissolve into the absorber. The ensuing GaOx layer provides a passivation effect leading to an increase of 80 meV in quasi-Fermi level splitting (QFLS) in comparison to the $CuInSe_2$ deposited directly on molybdenum, and *FF* above 71%[29] for 1.00 eV bottom cells. However, the production of industrial size modules is considerably easier when employing a sputter process. Consequently, sputter-deposited $InO_x$ has been proven to work in the same HTL system for sub-micron $Cu(In,Ga)Se_2$ films, with Rb post-deposition treatment, achieving PCE and *FF* above 18% and 77%, respectively.[36]

These results have been surprising, as it was previously shown, that $GaO_x$ layers block the current and significantly decrease the *FF* in solar cells with transparent back contacts.[37–40] It is therefore important to understand which factors contribute to $GaO_x$ layer being conductive or not. Understanding this question will have impact not only for hole transport layers, but also for semi-transparent and bifacial solar cells as well as for the top cells in stacked tandem cells.[41]

Ga gradients are usually employed in these low gap devices to passivate the back or the front interface.[15,16,42,43] Here, we investigate pure $CuInSe_2$ with 1.00 eV bandgap for tandem application, without Ga in the absorber. We specifically study the transport properties of $GaO_x$ in the HTL layer, which forms after growth, using sputter deposition for the oxide layer, compatible with commercial upscaling. We show, using absolute photoluminescence (PL) spectroscopy[44,45] that the Cu annealing stage is not needed for the passivation of the back contact. Electrical measurements show that this stage is not even required for the oxide layer to be conductive, contrary to what was found in the solution-processed oxide layer.[46] Furthermore, we explore the role of Na in the HTL. In most $Cu(InGa)Se_2$ solar cells Na diffuses from the soda-lime glass (SLG) substrate and was found necessary to obtain efficient absorbers.[47] Unlike in previously studied bifacial solar cells,[48] here we reveal that sodium doping does not play a crucial role in hole transport within $GaO_x$. We demonstrate that sodium incorporation, although needed to improve the quality of the bulk of the absorber, is not necessary for good hole transport properties inside the HTL. This has been proven by conducting experiments with and without $SiO_x$[48] as a Na blocking layer on top of the SLG substrate. These results are important not only for developing efficient $CuInSe_2$ bottom cells, but for further improving chalcopyrite bifacial devices or high bandgap top cells, since on transparent back contacts the formation of $GaO_x$ is thermodynamically favourable.[49] Additional investigation reveals that one of the limiting factors in the efficiency of these back-passivated bottom cells is a relatively high electrical diode factor, which tends to be much higher than the optical diode factor[50,51] of the absorber/buffer stack. This difference indicates dominant space-charge region (SCR) recombination as the main issue of the final devices. The best 1.00 eV device, with the finished cell structured as Glass/Mo/HTL/CISe/CdS/i-ZnO/AZO/$MgF_2$ (Ag alloying and heavy alkali treatment were not implemented in any of the samples), has shown a new certified $V_{OC}$ record[22,23] for $CuInSe_2$ without Ga of 552 mV and notable values of $FF$ = 69.9%, $J_{SC}$ = 42.7 mA/cm$^2$ (active area), η = 16.4% (active area).

## 2. Results

### 2.1 Cu doping

The samples in this study are prepared following the sequence of deposition shown in **Figure 1a**: on top of 500 nm sputtered molybdenum on SLG substrate, the two layers of the HTL (blue box) are deposited. First, a layer of 120 nm $CuGaSe_2$ (CGSe) is grown via co-evaporation, followed by 30 nm of $InO_x$ deposited by radio frequence (RF) sputtering. In the end, a standard $CuInSe_2$ three-stage process (grey box) is performed. An intermediate step (red box) involves annealing at 500 °C after the deposition of 20 nm to 30 nm of Cu precursor (which was necessary for the solution-processed HTL[32,46]) for samples "Cu HT", differentiating them from samples "HT", annealed at 500 °C without additional Cu, and samples "LT", without any high temperature annealing or additional Cu.

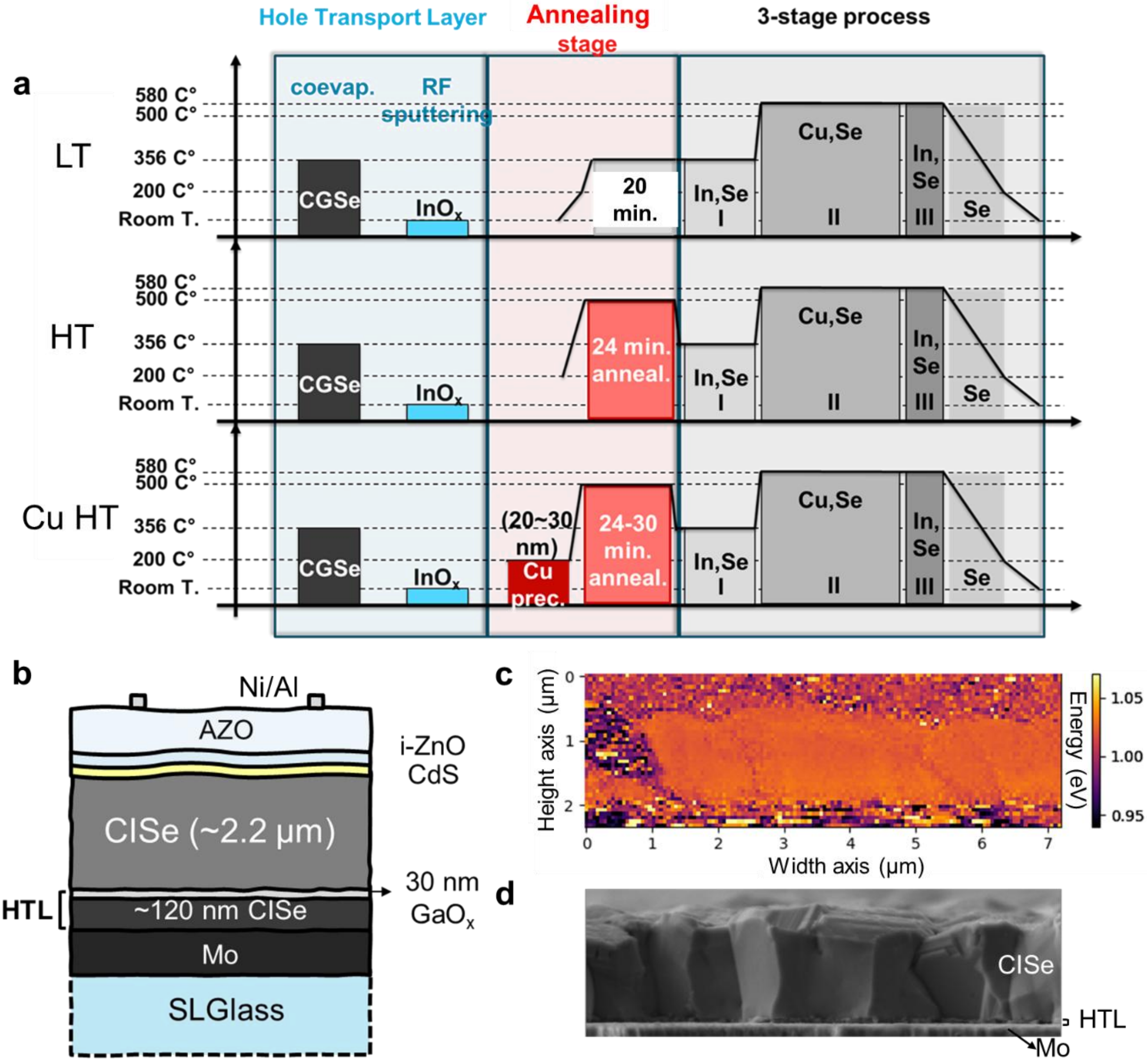

**Figure 1**. a) the three different processes for the CISe samples with HTL presented in this study: "LT" without Cu precursor or high temperature annealing, "HT" with only high temperature annealing, "Cu HT" with Cu precursor and high temperature annealing; b) structure of the finished device; c) cross-sectional CL map of the SLG/Mo/HTL/CISe stack: colour coded is the energy of the main luminescence peak ; d) cross-section SEM image of the same area as above, the scale is the same as in Figure c.

The three different processes all result in Cu-poor $CuInSe_2$ absorber layers, 2.2 μm thick, on top of the of two layers composing the HTL, as discussed later. The devices are finished by a CdS buffer layer by chemical bath deposition and sputtered i-ZnO and ZnO:Al layers. Then, Ni/Al grids are e-beam-evaporated, resulting in a structure as shown in **Figure 1b**. The absorber, grown in this way, has a homogeneous band gap, as demonstrated in the cathodoluminescence (CL) map of the cross-section (**Figure 1c**) (compare a similar absorber with Ga gradient in SI **Figure S1**). CL has been shown to be useful to detect composition and bandgap variations throughout the thickness of chalcopyrite absorbers.[52] This finding confirms that no significant amount of Ga diffused from the HTL into the absorber. In **Figure 1d**, the scanning electron microscopy (SEM) image of the exact same cross-section area demonstrates the presence of the HTL after the high-temperature deposition, as already found previously .[29] The sample presented in Figure 1c and Figure 1d has undergone to the process "Cu HT" (Figure 1a); the same structure is also observed in samples "LT", without Cu annealing stage (**Figure S2**). In **Figure S3a** and **Figure S3b**, the morphology, measured by confocal microscopy, of the absorbers with HTL ("LT" has similar results shown on "Cu HT" samples) is reported with its roughness value (with the average "Sa" value around 50 nm and root mean square "Sq" around 70 nm). Reduced roughness of chalcopyrite bottom cells has been reported to be crucial for the perovskite top cells growth[4,53], reducing the shunts, although efficient monolithic perovskite/CIGS cells have also been obtained with sputtered bottom CIGS cells with relatively high roughness.[54] Our absorbers on HTL show a roughness within the range of optimised co-evaporated chalcopyrites[54] and significantly lower than newly published values of sputtered absorbers, the latter evaluated with the same microscopy technique.[55] An interesting feature of the HTL, is that the examined samples exhibit lower surface roughness compared to those deposited directly on molybdenum (**Figure S3c**), suggesting an additional potential benefit of this back contact structure in tandem devices.

For the deposition of the HTL, first a Cu-poor $CuGaSe_2$ (EDS spectrum **Figure S4**) layer of thickness about 120 nm was co-evaporated at a substrate temperature of 356 °C. We emphasize that the sub-stoichiometry of this layer is essential for studying the influence of Cu on the oxide layer. In fact, in initial studies concerning this HTL, where the oxide layer was deposited via solution, its hole conductivity was possible only if the $CuGaSe_2$ layer was Cu-rich or a Cu precursor was deposited and annealed before the absorber growth,[29,32,36] as in the process "Cu HT" (Figure 1a). Afterwards, on top of the $CuGaSe_2$, $InO_x$ was sputtered at room temperature. After the absorber growth, the two layers of the HTL maintain the same thickness as deposited (**Figure 2a**) and show complete ion-exchange between Ga and In, as displayed in **Figure 2c** from STEM EDS mapping, resulting in $GaO_x$ on top of a $CuInSe_2$ thin layer, as previously found for solution deposited $InO_x$ (STEM-EDS-X line scan, in **Figure S5**).[29] The ion-exchange is also confirmed by SIMS measurements (**Figure S10**). Copper is distributed in both chalcopyrite layers, but it is below the TEM detection threshold in the $GaO_x$ phase. An enhanced selenium content is present at the back of the thin $CuInSe_2$ layer, close to the molybdenum back contact (bottom), indicating the presence of the $MoSe_2$ phase, which normally forms an ohmic contact in efficient CIGS cells.[56,57]

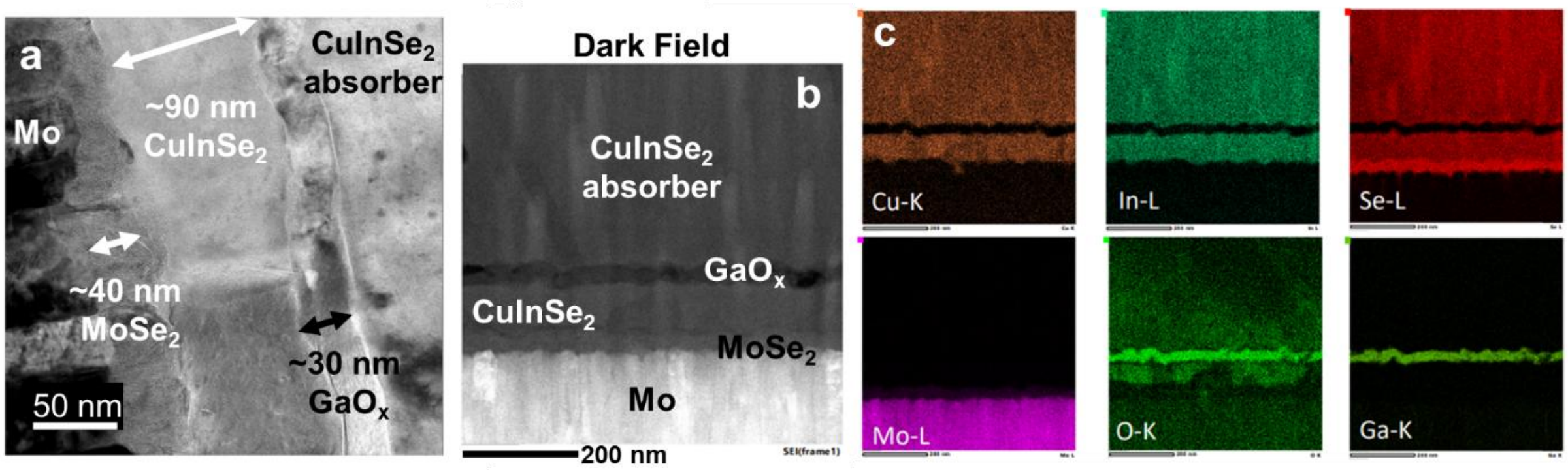

**Figure 2**. a) TEM cross-section image from Mo (left) to the $CuInSe_2$ absorber (right) on "LT" sample; b) dark-field STEM image of a cross-sectional image near the backside (Mo at the bottom) and ; c) corresponding elemental mapping measured by EDS showing the distribution of Cu, In, Se, Mo, O and Ga.

The $GaO_x$ layer, which has a large bandgap (around 4.7 eV) with substantial conduction band and valence band spike (electron affinity of 3.5 eV[58,59]) may hinder majority carrier transport from the absorber, which has a bandgap of 1.0 eV and an electron affinity of about 4.2 eV.[60–62] Thus, the valence band offset is several eV. A model of Cu related defects within the $GaO_x$, that align energetically with the valence of $Cu(InGa)Se_2$ has been proposed to explain the observed conductivity of $GaO_x$.[29,63–67] To understand the role of Cu in the oxide layer, we decided to grow the samples in the three different processes displayed in Figure 1a. These processes are distinguished by different amounts of Cu supply and heat. Since the hole conductivity in $GaO_x$ has been linked to Cu-related defects, we would like to see if Cu influences the transport properties of the $GaO_x$ layer. [32–36] In process "LT", the Cu annealing stage (red box), previously shown to be crucial for the conductivity of the solution-processed nonlow temperature stage of 20 minutes at 356 °C. In process "HT", this step was reduced to an annealing stage of about 24 minutes at 500 °C prior to the three-stage deposition. This procedure was introduced to investigate the possibility of interaction between the oxide and thin chalcopyrite layer due to the elevated temperatures. Meanwhile, the process "Cu HT" was the standard process previously developed[32,46], in which a layer of about 30 nm to 40 nm of Cu was deposited at 200 °C on top of the $InO_x$ and then annealed at 500 °C for a time that varied from to 24 minutes to 30 minutes, although the annealing time had no effect on the final result. This stage was thought to Cu dope the oxide layer, to create a deep defect band inside the material, allowing holes to reach the back contact while still maintaining the passivation effect for the electrons.

All three processes produce the same optoelectronic quality of absorber and passivate the back contact. **Figure 3a** displays absolute photoluminescence (PL) spectra of the various absorbers, covered by CdS to prevent oxidation.[68,69] All of them have a similar spectrum in terms of shape and photoluminescence quantum yield. The PL peak is at (990 ± 5) meV. All absorbers were produced with Cu-poor composition, but they exhibit slightly different Cu/In ratio (**Figure S6**). It should be noted that the Cu/In ratio in the three-stage process is determined by the duration of the third stage after the second stoichiometric point and is independent of the presence of the Cu-precursor. The PL peak shifts depending on the Cu content. It has been reported that decreasing the Cu content leads to a decrease in the bandgap as long as the composition stays within the chalcopyrite region in the phase diagram,[70,71] a phenomenon attributed to the interplay between Cu vacancies and anion displacements. The shape of the absorptance onset, which enters Planck's generalized law (see **Equation 1** in methods) describing the PL emission of a grey body in non-equilibrium,[45,72] can also shift the PL peak. An increase in the absorptance broadening σ (**Equation 4**) enhances the black-body contribution to the PL spectra, which is more pronounced at lower energies. Both effects can contribute to a shift of the PL maximum. It is noteworthy that, in the low energy wing of the spectra, some of the absorbers show increased emission below 0.9 eV, indicating the presence of a defect, detectable at room temperature. This defect can be attributed to the known 0.8 eV defect in chalcopyrites.[73] Even among absorbers of the same process type, the strength of the defect emission varies. Interestingly, a higher PL emission correlates with a lower emission below 0.9 eV., QFLS (**Figure 3b**), which is the maximum $V_{OC}$ achievable with the given absorber,[44,45] is obtained by fitting the high-energy slope of the PL spectra with **Equation 2**. They are reported as function of the absorptance onset broadening σ (Equation 4). The absorptance spectra A(E) all over the energy range is extracted from the PL spectra, after fitting the QFLSs with the assumption that $A$=1 at high energies. After, the first derivatives of the A(E) spectra are fitted with a Gaussian fit (Equation 4) to get their broadening σ. The broadening σ is related to the radiative loss in QFLS/voltage.[28,74–76] No correlation between QFLSs and broadening values is noted, indicating that the main loss mechanism is non-radiative. Higher σ values can also give absorber with QFLSs above 540 meV. However, the difference (**Figure S7)** between the bandgap energy ($E_g$), extracted from the peak of Gaussian fit of the first derivative of A(E) (Equation 4), and the radiative QFLS ($\Delta\mu^{rad}$, defined in **Equation 8**) of the absorbers grown with the three different processes shows a trend with σ, the absorptance broadening, indicating the expected increase in radiative loss with absorptance edge broadening. It is interesting to note that the decrease in the radiative loss with decreasing σ follows the same trend for all three types of samples. This behaviour is expected because a larger absorptance broadening σ leads to increased black-body emission and thus a decrease in the maximum achievable open-circuit voltage in equilibrium. It is also remarkable that, for all types of absorbers, the QFLS is particularly high compared to ungraded $CuInSe_2$ solar cells in the literature.[15,22,23] Urbach energies, indicative of the static and dynamic disorder,[77,78] are not fitted in this work, because the presence of the previously mentioned defect at lower energies makes any exponential fit to the absorption onset unreliable. In **Figure 3c**, we correlate $V_{OC}$s and *FF*s of finished devices, deposited via the three different processes. The observed *FF*s confirm that a Cu precursor stage is not needed for making the oxide conductive. Indeed, Cu-annealed devices ("Cu HT") show lower *FF*s than those without Cu annealing ("LT"). Interestingly, $V_{OC}$s of the best devices are comparable to the best back-graded cells without heavy alkali treatment or Ag alloying in the literature [15,79] and there is an improvement in $V_{OC}$ of nearly 50 mV compared to the first device with solution-processed HTL published,[29] demonstrating the high potential of this passivating structure. **Figure 3d** displays *J*-*V* characteristics of the best devices with the three different processes. No differences in short-circuit current are noticed as shown in detail in **Figure S8**.

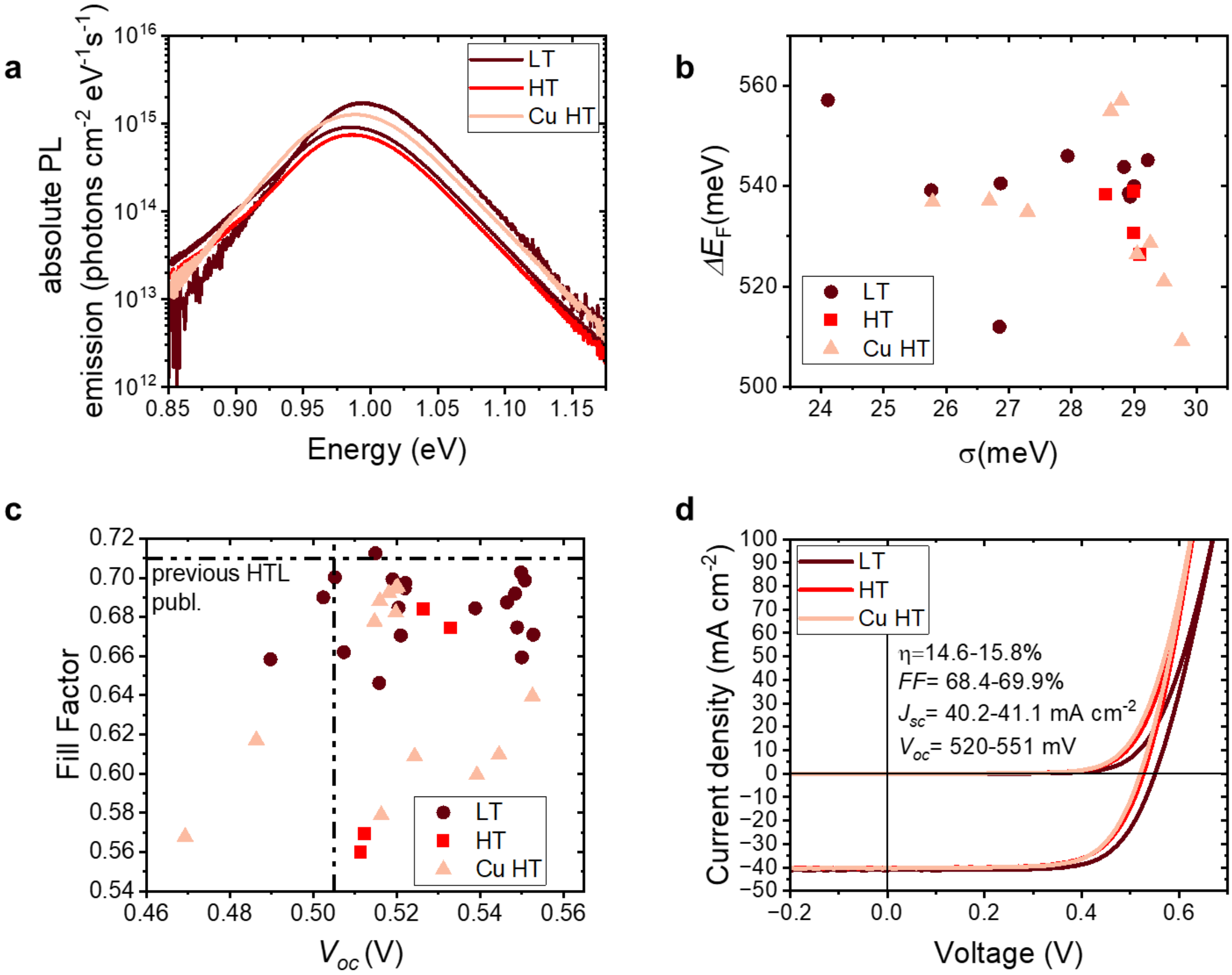

**Figure 3**. a) Room temperature absolute photoluminescence (PL) of different Mo/HTL/absorber/CdS stacks. There are two different measurements of two different samples from the "LT" process to demonstrate the variability from the same process; b) Quasi-Fermi level splittings (QFLS) and absorptance broadening σ from PL of Mo/HTL/absorber/CdS stacks; c) *FF*s and $V_{OC}$s of the finished devices. The dash-dotted lines show $V_{OC}$ and *FF* of the best device, previously reported for $CuInSe_2$ on HTL.[29] d) *J-V* curves of the best devices produced with the three different processes.

The series resistances extracted from one-diode fits of the *J-V* characteristics of the samples grown on HTLs with the three different processes are reported in **Figure 4**, together with the FFs and the electrical diode factors (EDF) of all the devices. All of them have values ≤ 0.65 $\Omega\cdot cm^{-2}$, showing that $GaO_x$ layer is hardly adding additional series resistance to the device. It is noticeable that reducing the series resistance and achieving a low EDF is essential to increase the *FF*s and consequently, the overall efficiency. The dependence on the EDF appears more systematic than the dependence on the series resistance. However, even in the best performing cells, the EDF does not fall below 1.67. The "LT" samples exhibit the lowest and the highest series resistance, while the "Cu HT" samples have higher series resistance than many of the "LT" samples, confirming again that the additional Cu is not needed to make the $GaO_x$ conductive. Moreover, the two highest series resistance samples have the second highest *FF*s, clearly suggesting that the EDFs play a more important role in the *FF*s.

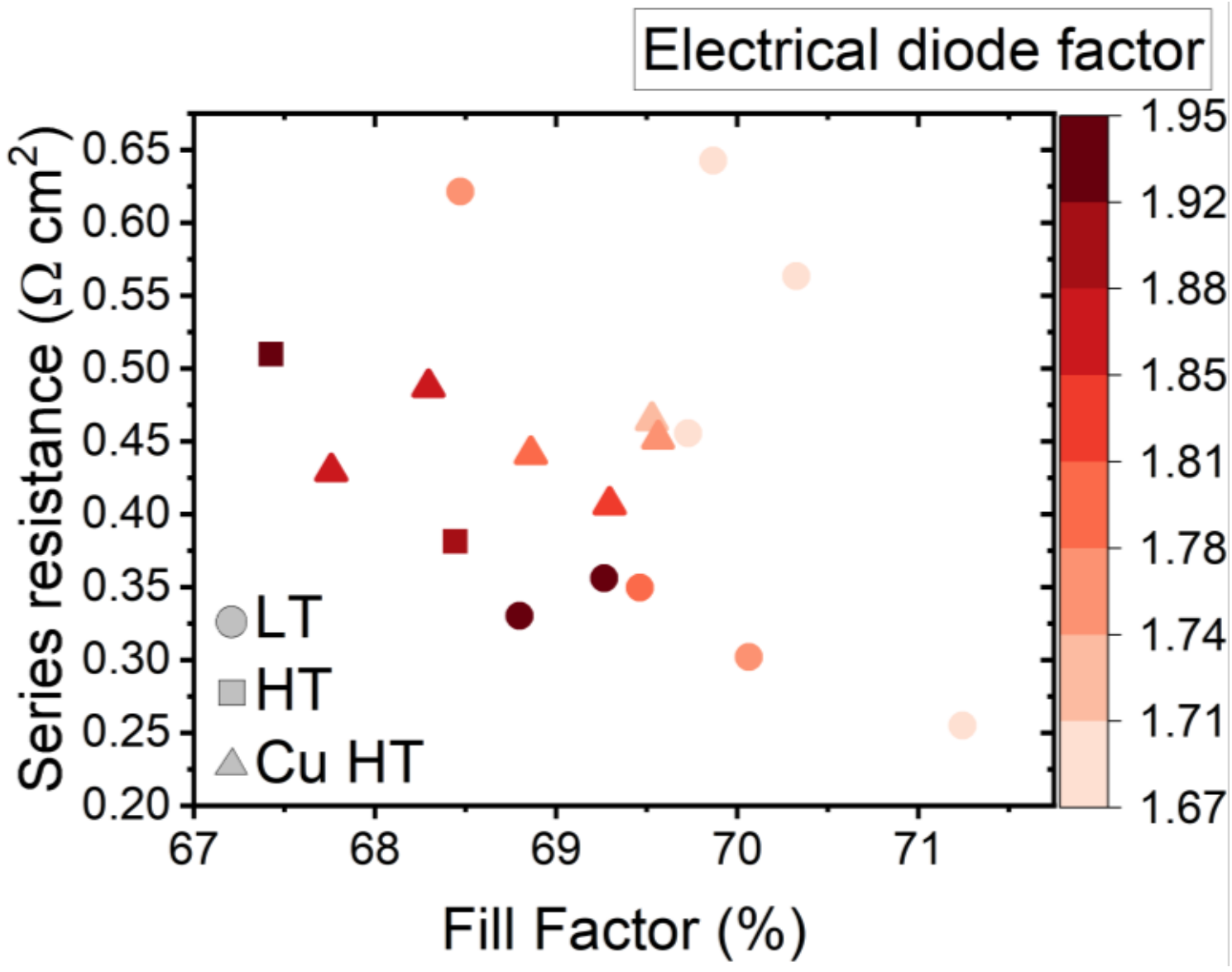

**Figure 4**. *FF*s plotted against series resistance and EDFs of the devices produce with the three different processes: "LT" (circles), "HT" (squares) and "Cu HT" (triangles).

It is worth noting that the Cu precursor is needed to prevent the formation of an (uncontrolled) Ga gradient, if standard $Cu(In,Ga)Se_2$ are grown on top of the HTL.[36] But its presence is not necessary to make the $GaO_x$ layer conductive. However, we cannot exclude that Cu, which is extremely mobile, may still diffuse into the oxide during absorber or during the oxide growth to create deep defects that allow the hole transport through the $GaO_x$ layer to the back contact. Especially considering that sputtering is a high energy process, already the two layers of HTL might interact with each other. Also, Cu diffusion during the absorber deposition process is possible, which may have happened to a lesser extent in the solution-processed HTLs because of lower purity of the $InO_x$/$GaO_x$ films.[29]

### 2.2 Na doping

Bifacial chalcopyrite cells are grown on transparent conductive oxides as back contact.[39,80] The formation of $GaO_x$ is thermodynamically favourable in such cells with indium and zinc based back contacts[49,81,82]. The unintentional GaOx layer has been reported to block the current.[48] Therefore, studying the transport properties of this layer is also crucial for the development of efficient solar cells with transparent back contact, whether for single junctions or top cells in tandem applications. In fact, SnO:F (FTO), where $GaO_x$ presence is not predicted, manifested other drawbacks such as potential fluorine loss, increasing the sheet resistance of the back contact,[83] and lower transparency than tin/hydrogen-doped $In_2O_3$ (ITO/IOH) and aluminum-doped ZnO (AZO).[84]

It has been documented that sodium can diffuse into the $GaO_x$ layer in bifacial $Cu(In,Ga)Se_2$ solar cells, a phenomenon facilitated by the presence of Ag.[85] Trap-assisted tunnelling due to sodium defect generation in $GaO_x$ at the interface of ITO/CIGS of bifacial cells has been previously proposed and it was found that the blocking behaviour in the devices was strongly dependent on Na presence during the formation of the high bandgap oxide phase.[48] On the other hand, sodium has been reported to promote the $GaO_x$ formation at the interface between the absorber and the transparent back contact.[37,49,86] To better understand Na doping inside our $GaO_x$ HTL, we study $CuInSe_2$ films on HTL and on Mo without HTL with and without additional Na supply. To this end, we deposited sample "LT" (Figure 1a) on top of the HTL structure or directly on molybdenum, with and without a $SiO_x$[48] barrier between the substrate and the metallic back contact, which prevents the sodium diffusion from the soda lime glass (SLG).

In **Figure 5a**, the absolute photoluminescence (PL) spectra of the four CdS-covered $CuInSe_2$ absorbers implies that sodium is required to enhance the bulk quality.[87,88] Both samples grown without Na blocking barrier, whether on the HTL or on Mo, exhibit higher PL yield and so the lowest non-radiative losses.[78] Sufficient Na quantity is introduced via diffusion from the glass substrate. Additional Na precursors in HTL samples, deposited on top of the oxide layer of the HTL or on the molybdenum back contact, lead to a degradation of the absorber, resulting in a reduced PL yield (**Figure S9**). This finding suggests that sufficient sodium can diffuse partially through the oxide layer, although SIMS measurements show that the HTL blocks part of the Na diffusing from the substrate (**Figure S10**). It is also worth noting that even for the sample with a Na barrier, a non-zero Na count rate was found within the various using SIMS (~ half in the absorber layer and ~ 1 order of magnitude less in the GaOx and Mo) (**Figure S11**). This may partly be due to the contamination from the growth chamber and from $Na_2Se$ volatile species from the SLG substrates during the growth.[89] Moreover, the smaller grains in the HTL Na barrier sample (**Figure S12**) lead to a higher Na signal in SIMS measurements. This is caused by the predominant Na segregation at grain boundaries.[90,91] Therefore, a higher fraction of grain boundaries results in an increased Na detection in SIMS profiles. Therefore, the SIMS signal overestimates the real intragrain Na concentration, compared to sample HTL without Na barrier. The samples grown with sodium barrier display lower PL yield and a visible defect in the lower energy wing of the spectra, likely attributable to the known 0.8 eV defect, which has a broad density of states (grey box in Figure 5a).[73] The best sample is the $CuInSe_2$ with HTL and Na supply, with an enhancement in QFLS of about 70 meV compared to the Mo sample with Na, as also reflected in the $V_{OC}$ improvement in *J-V* characteristics of about 60 mV in **Figure 5c**, and a reduction in tail states (**Figure S13**). In contrast, samples deposited with a sodium barrier show no difference between the back passivated sample and the one grown on molybdenum. These samples are limited by the poor bulk quality and the reduction in non-radiative recombination at the back contact does not significantly affect the PL yields and $V_{OC}$s. TEM investigation of the sample on HTL with Na barrier (**Figure S14**) shows that the ion-exchange and HTL structure remain the same as in the standard HTL sample (Figure 2). In **Figure 5b**, the external quantum efficiency (EQE) indicates that there is no difference between the HTL samples, with and without sodium, and the Mo with Na sample. There is a reduction in the Mo Na barrier sample in collection at the longer wavelengths, where the photons are absorbed close to the back contact and the back side passivation is crucial. In this case, the poor bulk quality hinders carrier mobility, together with the lack of back passivation, reducing the collection of photogenerated charged carriers. The HTL samples exhibit a steeper EQE edge ($\sigma$ around 8 meV lower as reported in **Figure S15b**) than the Mo ones. Mo samples show a red shift of about 10 meV in the EQE onset, as also confirmed in the A(E) extracted from PL (**Figure S15a**). A small difference in Cu/In ratio[70] (**Figure S16**) is present between the HTL and Mo samples.

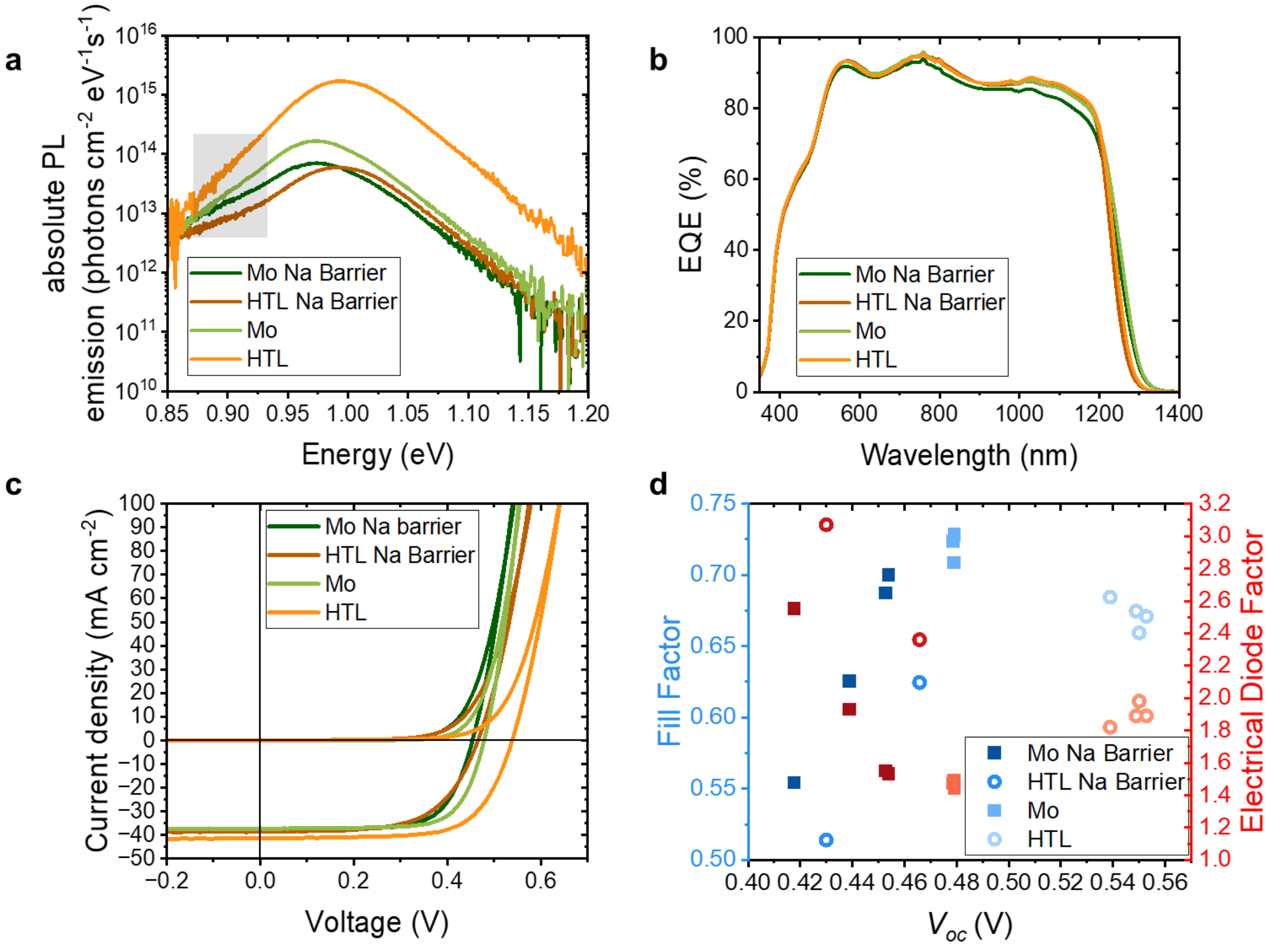

**Figure 5**. a) Room temperature absolute photoluminescence (PL). The lower energy defect disappears with Na; b) External quantum efficiencies of the finished devices; c) *J-V* characteristics of the best finished devices; d) *FF*s (blue), EDFs (red) and $V_{OC}$s of the finished devices

In the *J-V* plots all samples display a diode behaviour, with no S-shaped curves or indication of a barrier, in contrast to what was previously reported when sodium is absent in the $GaO_x$ layer.[48] $V_{OC}$s follow the same trends as PL yields, with the best-performing sample being the HTL with Na doping. In **Figure 5d**, $V_{OC}$s of all devices are plotted against *FF*s of the same cells (blue), along with the electrical diode factor (EDF) extracted from a one-diode fit of light *J-V*s(red). It is noteworthy that, although *FF*s depend on $V_{OC}$s, in the case where nothing else changes, the trend is not observed here: the highest $V_{OC}$s are not associated with the highest *FF*s. Indeed, the highest *FF*s are found in the samples on Mo, compared to the HTL counterparts, which show the higher $V_{OC}$ because they passivate the back contact. The *FF*s are also determined by both series and shunt resistance and the EDFs. An increase in EDF leads to a decrease in *FF*. This is not caused by a high series-resistance in the samples, as explained later, but it is due to elevated EDFs. This suggests that *FF*s are mainly EDF-limited, with an increment of the latter leading to a decrease in *FF*s. In chalcopyrite technology, EDF is always greater than 1, due to metastable defects such as copper-selenium divacancies.[92–94] These defects increase the net acceptor density upon minority-carrier injection and, consequently, shift the majority quasi-Fermi level toward the valence band upon illumination, even in the low injection case. Here, the *FF*s of HTL samples are lower than those on Mo. The increase in EDFs from Mo to HTL samples can be explained by a lower recombination in the back side: the additional non-radiative recombination in the Mo samples of the diffused minorities towards the back contact shifts back the EDFs towards 1.[24] The decrease in EDFs in the sodium-containing-samples may be explained by a higher doping level inside the absorber and thus to a larger concentration of holes,[24] which makes the shift of the effect of the metastable defects less impactful, as well as to a lower number of 0.8 eV defects, which can cause more space-charge region recombination, as explained later.[95]

Sodium doping does not play a major role in hole transport within $GaO_x$, unlike previous results in bifacial solar cells.[48] However, it is remarkable to note that the high-bandgap oxide layer we investigated via HRTEM (**Figure 6a**) is partially crystalline, as can be seen by the clear pointlike and well ordered diffraction peaks in the Fast Fourier transformed of an area of the the HRTEM image of the oxide layer (**Figure 6b**). Partial crystallinity has been observed with or without Na from the SLG substrate (**Figure S17**). [29] In all previous reports on bifacial cells and the formation of $GaO_x$ at the interface with the transparent back contact, the layer has always been shown to be amorphous[39,48,49]. The nano-crystallinity we observe could indeed be significant for charge transport. Interestingly, the 30 nm $InO_x$ layer deposited via sputtering is initially amorphous (**Figure S18**). However, during absorber deposition, simply due to the temperature, it begins to transform into $GaO_x$ and becomes crystalline (**Figure S18**). This is visible in the shift of the chalcopyrite peak to lower angle (i.e. from $CuGaSe_2$ towards $CuInSe_2$)[96] and the appearance of the oxide peak, shifted towards higher angle (i.e. form $In_2O_3$ towards $Ga_2O_3$) [97,98], which also become broader[99]. The start of the ion exchange between the two layers, caused by the elevated temperature, is also confirmed by Raman spectroscopy (**Figure S19**).

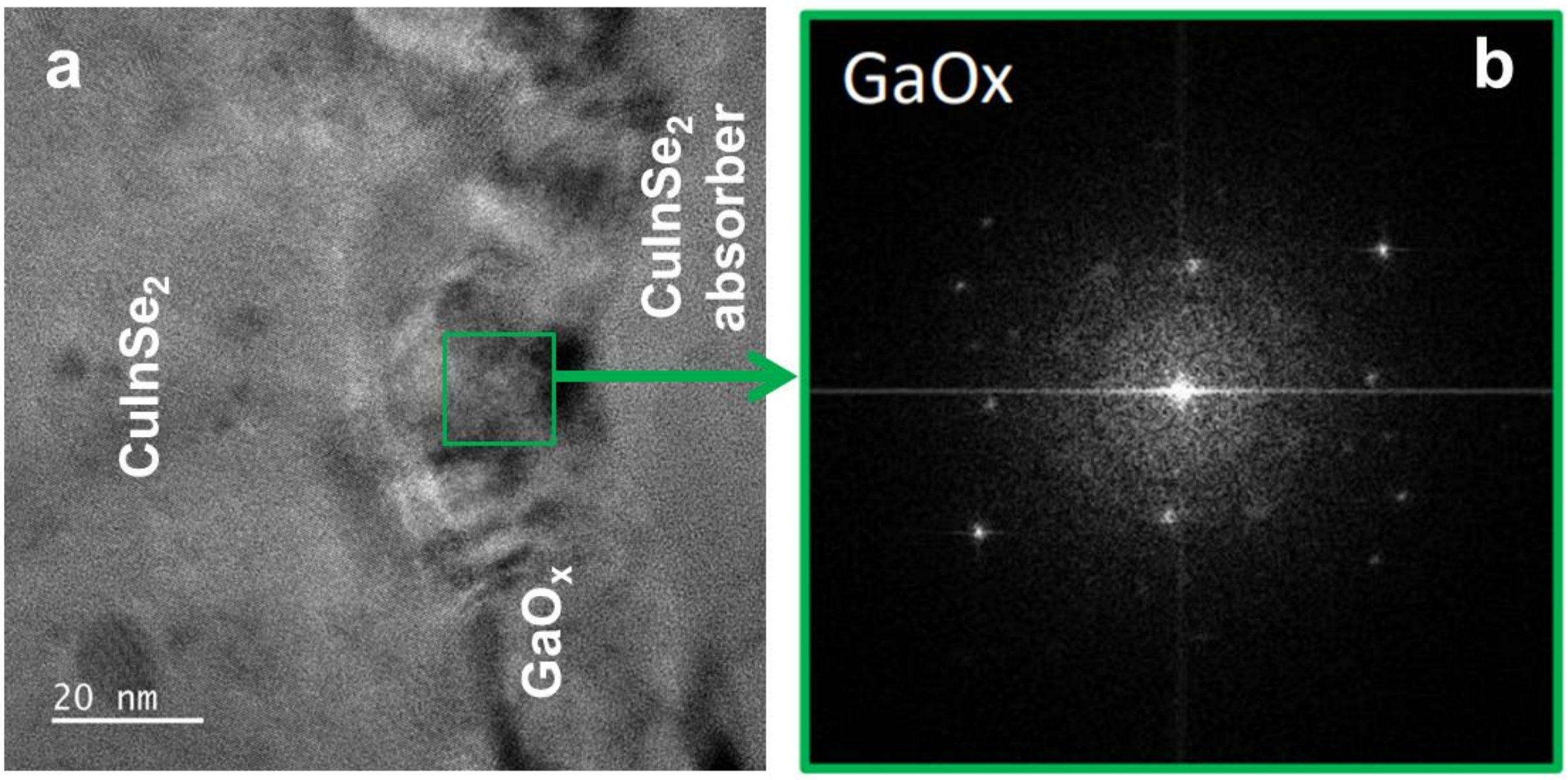

**Figure 6**. a) High-resolution TEM images of the back-side region; b) Fast Fourier transform (FFT) of an area (green square region) in the $GaO_x$.

### 2.3 Oxygen flow

It has been reported that the conductivity of $Ga_2O_3$ depends critically on the oxygen deficit.[100,101] The exact composition of the $GaO_x$ hole transport layer is unknown; however, we attempted to modify its composition via varying the oxygen flow during the $InO_x$ sputtering process. The *J-V* characteristics of the final devices fabricated under different oxygen flow conditions (0, 0.5, 1, 10 sccm $O_2$) during $InO_x$ deposition are presented in **Figure 7**. The effect of passivation can be concluded from the open-circuit voltage and the *FF* gives hints for the conductivity of the $GaO_x$ HTL. No trend is observed between the oxygen flow rate and the passivation and transport effects of the $GaO_x$ hole transport layer. The HTL based on the $InO_x$ film grown with 1 sccm O2 during the sputter process shows the lowest $V_{OC}$ and *FF*. But higher oxygen flow recovers both. We therefore conclude that additional oxygen during the $InO_x$ sputter process does not influence the passivation and transport properties of the ensuing $GaO_x$ HTL. Further optimisation of the $InO_x$ layer can be found in the SI.

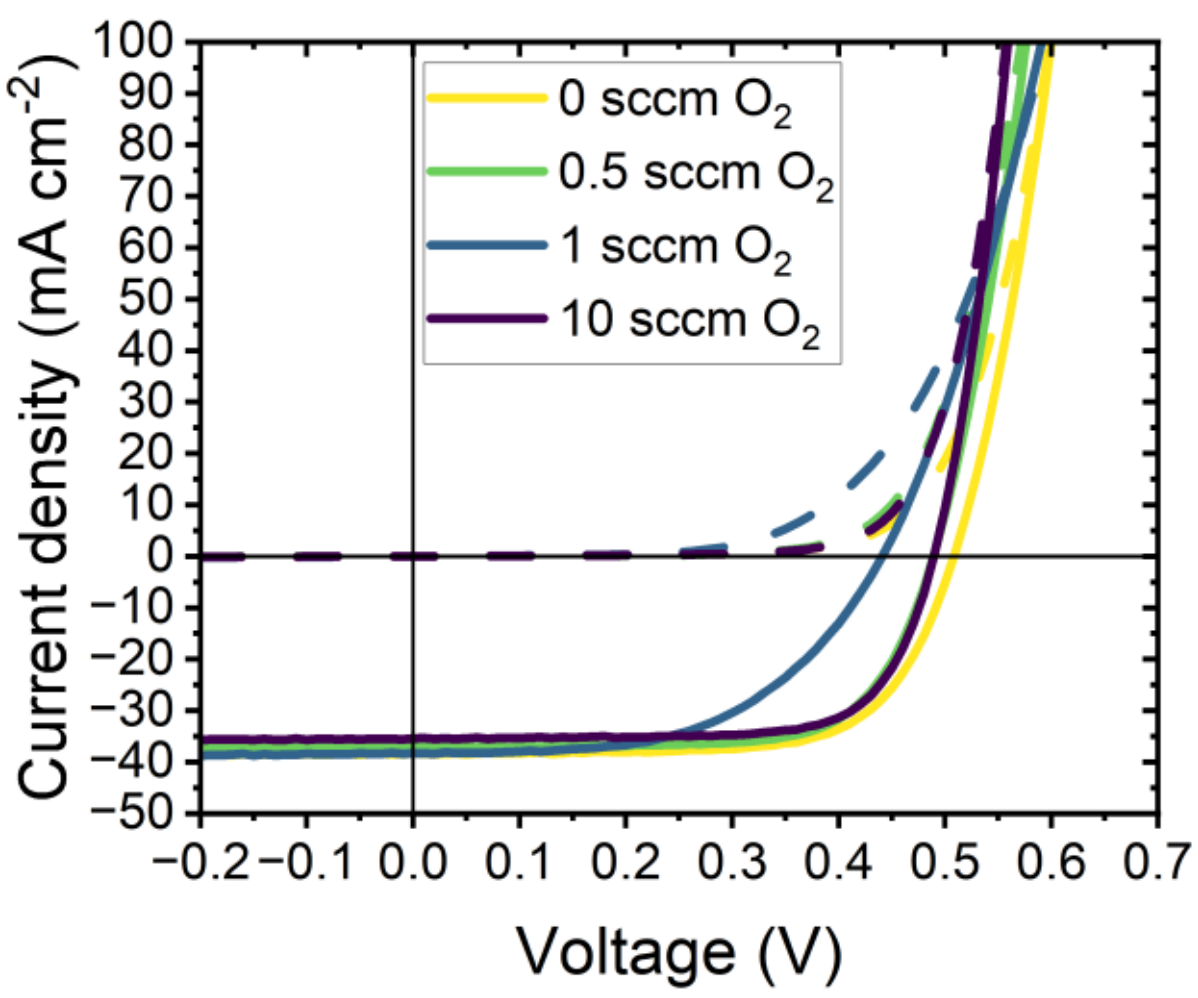

**Figure 7**. *J–V* characteristics of completed devices as a function of oxygen flow during $InO_x$ deposition.

### 2.4 Fill Factor Losses

As discussed above, one of the main performance limitations of the presented bottom cells is the fill factor (*FF*), which is constrained by a high electrical diode factor (EDF) and exploits only the 82% of its Shockley–Queisser limit for 1.0 eV bandgap solar cells.[2,102] The high EDFs cause the *FF*s to be about 6%(absolute) lower than the expected value for the same $V_{OC}$s with EDFs equal to 1. To better understand the underlying causes, optical diode factor (ODF)[50,51,92,93]measurements were performed. The ODF is the exponent in the power law that describes how photoluminescence (PL) emission varies under excitation-dependent measurements across several orders of magnitude. If band-to-band recombination is the only active recombination mechanism and no competing channels are present, the PL intensity increases linearly with excitation power, and the ODF equals a value of 1 (**Equation 9**).[103] The EDF will be identical to the ODF when the finished cell exhibits the same dominant recombination paths as the absorber alone.[50] If the cell finishing adds new recombination channels (e.g. recombination in the space charge region) the EDF is larger than the ODF.[92] Thus, the minimum EDF of the finished device can be predicted using a simple measurement on the absorber alone.[50,51] Moreover, in chalcopyrite materials, the ODF is always greater than 1 due to the presence of metastable defects.[92,93]

In **Figure 8a**, the ODF of several "LT" and one "Cu HT" absorber are shown; their value is consistent with those previously reported for chalcopyrite materials and for high-efficiency graded CIGSe bottom cells.[92] **Figure 8b** displays the ODF values for the samples from the Na doping series (see also Figure 5). The Mo sample exhibits the lowest ODF, which can be attributed to detrimental non-radiative recombination at the back side. The dominant minority carriers' recombination at the back interface reduces the ODF towards 1 and screens the real value of the absorber.[24] The samples with HTL exhibit higher ODF values, which can be attributed to a reduction in back surface recombination, previously estimated for these samples to be around $10^3$ cm·s$^{-1}$,[29] against a higher recombination velocity for the Mo samples about $10^6$ to $10^7$ cm·s$^{-1}$.[21] This reduction allows for a more accurate estimation of the ODF of the absorber itself. This holds true for the sample with sodium, whereas in the sample HTL with a Na barrier, additional bulk recombination masks the effect of metastabilities, resulting in a lower ODF. The Mo Na barrier sample also exhibits a low ODF, influenced by significant recombination at the back contact, although this ODF value is still higher than that of the Mo sample. This can be explained by a lower doping level, which reduces the majority carrier concentration and increases the absorber's sensitivity to the donor-to-acceptor transformation of metastable defects, but also by worse bulk properties, which prevents the minorities to reach the Mo back contact.

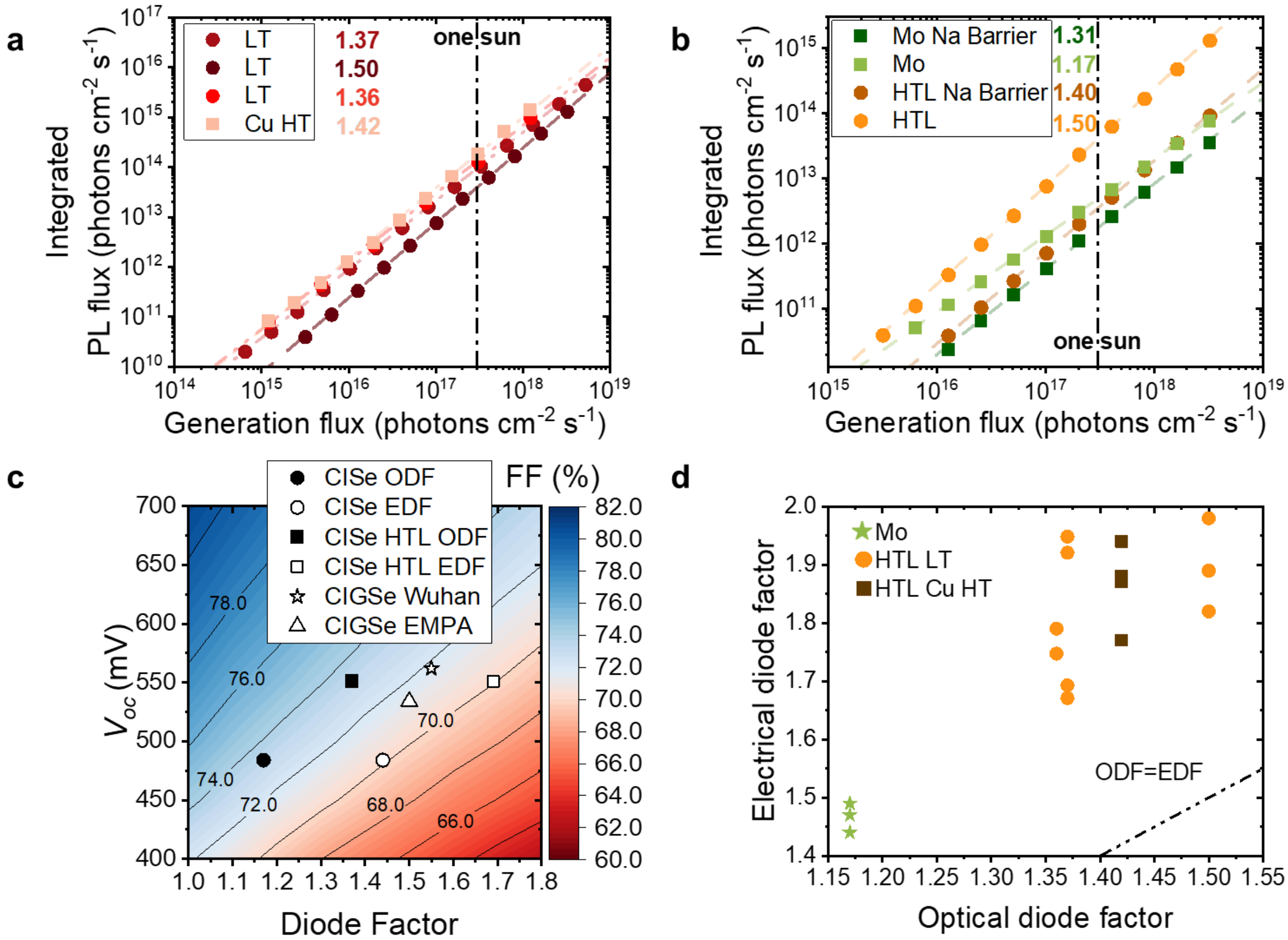

**Figure 8**. a) Generation flux-dependent PL flux of "LT" and "Cu HT" absorber/CdS stack; b) Generation flux-dependent PL flux of the absorber/CdS stacks in the Na; (the numbers in the legend in a) and b) give the extracted ODF). c) Simulation of *FF*s varying the $V_{OC}$s and the diode factor in one-diode model with the symbols representing the absorbers (filled) and the cells (empty). The CISe sample is the "Mo" sample in Figure 5, while the CISe HTL is the "LT" back-passivated sample giving the record certified device (Figure 3d). Highly efficient CIGSe bottom cells from Wuhan University[15] and EMPA[79], without Ag and Rb-PDT, are also reported for comparison; d) ODFs and EDFs of the studied absorbers and cells.

Interestingly, the EDFs of different cells from the three fabrication processes (Figure 4) are significantly higher than the ODF values measured on the bare absorbers (Figure 8a). A high EDF indicates a fundamental loss mechanism in these bottom cells, and this is also observed in other high efficiency samples from different laboratories.[104] **Figure 8c** presents a simulation of *FF*s, varying the $V_{OC}$, with the saturation current density ($J_o$), and the diode factor, for a cell with a current density of 42 mA·cm$^{-2}$, series resistance of 0.5 Ω·cm², and shunt resistance of 1000 Ω·cm². On the graph, we include our best HTL CISe sample (squares) ("LT" from Figure 1a) and Mo sample without back passivation (Mo from Figure 5) (circles). Filled symbols represent the ODF of the bare absorber, while empty symbols correspond to the EDF of the final device. Inserting the fitted EDF values (an example of the fit in **Figure S20**), from *J-V* characteristics for the measured $V_{OC}$s, the *FF*s of our finished devices are well represented in the simulation. However, it can be observed that, if the ODFs of the absorbers were maintained, a *FF* gain of approximately 3% could be achieved with the same $V_{OC}$s. Samples from Empa[79] (triangles) and Wuhan University[15] (stars) are also shown. The samples from these institutes are passivated with a Ga gradient toward the back side. They do not include Ag or Rb post-deposition treatment, nor Ga front gradients, to be comparable with our bottom cells, since all of those are known for improving the $V_{OC}$.[15,16,43,79] While they exhibit $V_{OC}$s values similar to our passivated sample without Ga, their *FF*s are slightly higher due to lower EDFs, although they are still above 1.5 for all reported samples. This highlights the high electrical diode factor as one of the main limitations currently affecting the efficiency of chalcopyrite bottom cells. In **Figure 8d**, the EDFs of the cells are plotted against the ODF values of their corresponding absorbers. A clear correlation between the two can be observed. However, they are still far from identity (the dashed line, which represents the ideal case where EDF equals ODF). This indicates that, when the p–n junction is formed, additional loss is introduced into the cell. The recombination current is dominated by the Shockley-Read-Hall (SRH) recombination in the space-charge region (SCR), where the high-injection condition leads to diode factors closer to 2.[104]

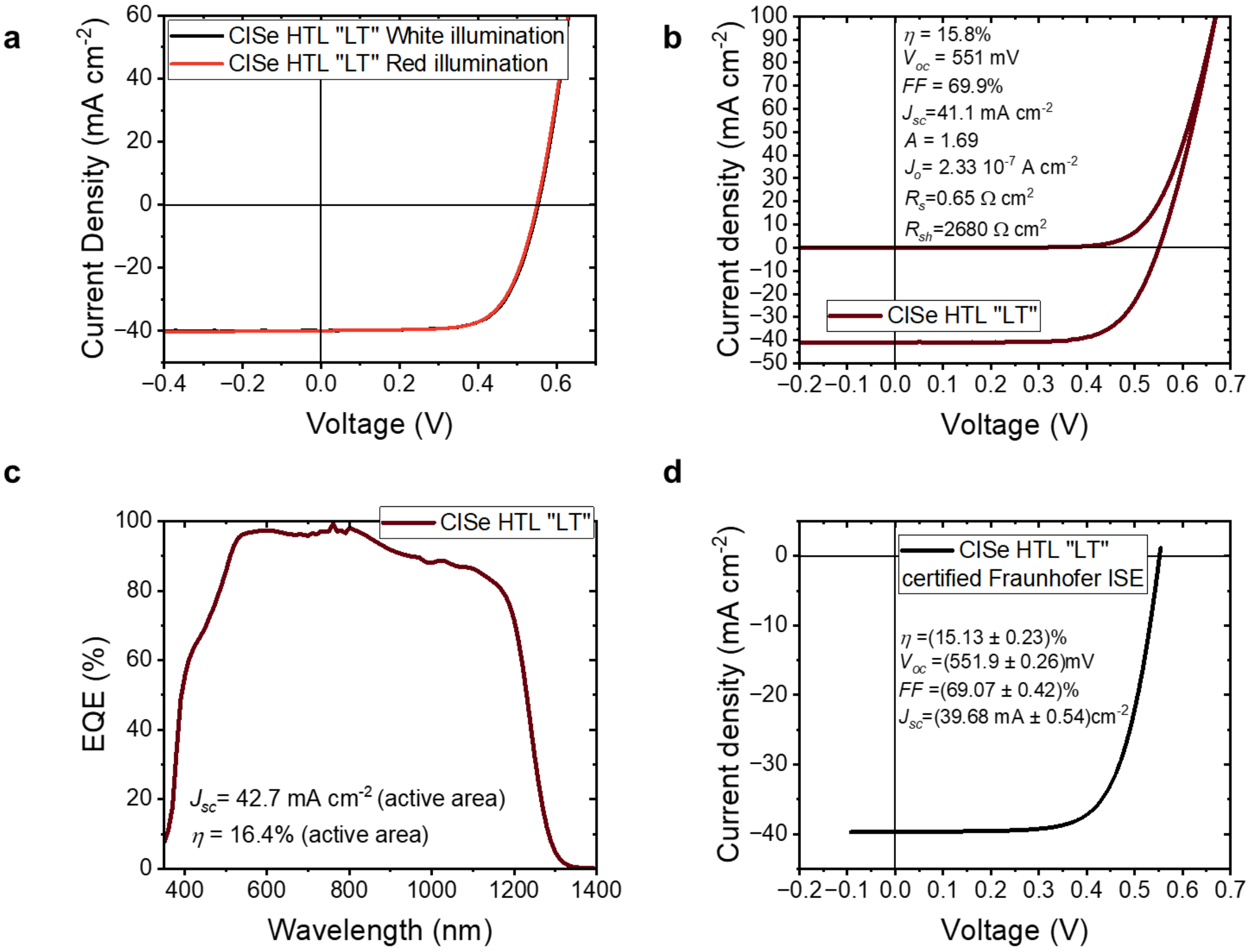

**Figure 9**. a) *J–V* characteristics under white and red illumination of CISe HTL sample ("LT") of Figure 5; b) *J–V* characteristics (in-house) in the dark and under illumination of the record "LT" CISe HTL sample; c) EQE spectrum (in-house) of the record "LT" CISe HTL sample; d) *J–V* characteristics, certified Fraunhofer ISE, of the record "LT" CISe HTL sample.

When these cells are used as bottom cells in a tandem device, they will not receive white light, but filtered light which misses the blue parts of the spectrum. This is relevant, because it has been reported that in CIGS cells,[105–107] the absence of high-energy photons can lead to changes in the band structure that may cause S-shaped *J-V* plots, although this is not a universally observed trend.[7] In **Figure 9a**, *J-V* characteristics of the sample "LT" is shown under both white (simulated AM 1.5) and red illumination (660 nm), adjusted to the same short circuit current. No differences are observed, which shows that these cells are suitable as bottom cells.[102,103][7] **Figure 9b** and **Figure 9c** show the *J-V* characteristics and the active area EQE spectrum (measured in-house) for our best device fabricated using the "LT" process. The efficiency displayed in Figure 8b, as well the *J-V* characteristics, are measured on the finished device with the grids on top. The EQE spectrum and the extracted short-circuit current and the efficiency in Figure 8c are obtained on from the active area of the device, without considering the shading of the grids. This shows the real potential of the cell in a 2-terminal tandem configuration, where the grids are not necessary. This device was produced without any heavy alkali or Ag addition. The finished cell demonstrates a record-certified $V_{OC}$ for a $CuInSe_2$ absorber[22,23], without Ga gradient, opening new possibilities for this bottom cell design. The record $V_{OC}$ is obtained from the *J-V* characteristics in **Figure 9d**, certified by Fraunhofer ISE (**Figure S21**). The EQE measurement (Figure 9c) clearly shows parasitic absorption by the CdS layer, but this is not considered critical if the cell is operated in a tandem configuration. The losses in the long-wavelength region could be improved by enhancing the bulk quality or by using a TCO layer with lower free-carrier absorption, such as bias-ZnO.[108]

## 3. Conclusions

In this work, we have demonstrated that an HTL, composed of 30 nm $GaO_x$, can provide an excellent passivation of the back contact. The $GaO_x$ HTL layer shows partial crystallinity. Optical and electrical measurements indicate that a copper annealing stage, prior the 3-stage absorber deposition, is not needed for the conductivity of the oxide layer. This is likely due to the different crystallinity and reduced impurity content in the sputtered layer compared the previously used solution-processed one, where Cu annealing was necessary. We have also found that the sodium presence, although necessary to improve the bulk quality of the absorber, does not play a crucial role in the conductivity of the high bandgap oxide. Indeed, samples passivated with the HTL, grown with $SiO_x$ as a Na blocking layer on top the SLG substrate, show good transport properties of the back contact, contrary to previous observations in bifacial devices. A main limitation of the studied bottom cells is its high diode factor, which decreases the *FF* of the finished devices. This trend appears to be common among efficient bottom cells fabricated in different laboratories. Through optical diode factor measurements we have understood that the diode factor worsens after TCO deposition and is not only caused by the absorber metastabilities, but due to dominating recombination in the space charge region. Based on our novel HTL, we have fabricated a device with 1.00 eV bandgap that has reached an active area PCE above 16%. The device has proven a record $V_{OC}$ of 552 mV for pure CISe bottom cells, as certified by Fraunhofer ISE. These improvements highlight the potential of CISe bottom cells with this HTL design.

## 4. Methods

### 4.1 Sample preparation

The finished cell is structured as Glass/Mo/HTL/CISe/CdS/i-ZnO/AZO/$MgF_2$, for the HTL samples, and as Glass/Mo/CISe/CdS/i-ZnO/AZO/$MgF_2$, for the Mo samples. The samples with Na barrier contain a 100 nm SiOx between the glass and the Mo. The 500 nm molybdenum was prepared via sputtering (**Figure S22**). The HTL is prepared as a double layer of $CuGaSe_2$ and $InO_x$. 120 nm $CuGaSe_2$ were deposited by co-evaporation in a VEECO Gen930 MBE system with a set substrate temperature of 356 °C. 30 nm $InO_x$ were sputtered in an AJA ATC Orion system using a 2 in $In_2O_3$ target, sputtered with 130 W power at 1 mTorr without additional substrate heating or bias voltage. The thickness was measured by ellipsometry. The obtained spectra were fitted from 0.8eV to 2.8 eV with Cauchy model. Additional information about this layer and its optimization is reported in **Figure S23**. The 2.2 µm $CuInSe_2$ absorber was deposited in the VEECO MBE system using three-stage co-evaporation process. In the first stage, In and Se were evaporated at a substrate temperature of 356 C°. During the second stage, Cu and Se were deposited with the substrate set at 580 C° and the samples reach a Cu-rich composition (Cu/In ratio of about 1.20, estimated from the time after the stoichiometric point, which was determined from the output power variation to keep the substrate temperature constant[109,110]). In the last stage, to finally produce Cu-poor samples, In and Se are supplied at 580 C° substrate temperature. The absorber layers were covered with approximately 30 nm of CdS via chemical bath deposition, after being chemically etched with 5% aqueous KCN solution for 30 s to remove potential residual oxides. To complete the devices, 100 nm of intrinsic ZnO (i-ZnO) and 300 nm of aluminum-doped ZnO (AZO) were deposited on the absorber/buffer stacks via RF sputtering in AJA system, followed by e-beam evaporation in an Univex system with Ferrotec e-beam of 2 µm Ni/Al metallic grids and 110 nm of $MgF_2$ anti-reflection coating.

### 4.2 Absolute photoluminescence

Absolute PL measurements were performed using in-house spectral and intensity calibration of the PL spectra.[111]All samples were excited with a continuous-wave laser (660 nm) of 2.2 mm diameter at a photon flux density equivalent to 1 sun. The PL signal was first collected by two parabolic mirrors and a 550 µm optical fiber to a monochromator and the detected by an InGaAs array detector. The quasi-Fermi level splittings (QFLSs) were extracted by fitting the high-energy tail of the PL spectrum, assuming an absorptance A(E) = 1 (this assumption may lead to a slight underestimation in QFLS by about 5 meV, as A(E) < 1 in this region)[28], using Planck's generalized law under the Boltzmann approximation at a fixed temperature of 296 K, as measured during the PL measurement.

$$\Phi_{PL}(E) = A(E)\Phi_{BB}(E)\, exp\left(\frac{\Delta\mu}{kT}\right) \tag{1}$$

All symbols are explained in Table 1. The high-energy range of the photoluminescence spectrum of a sample is fitted using the generalized Planck's law [44,72] at 296 K under the Boltzmann approximation.

$$ln\frac{h^3c^2\Phi_{PL}(E)}{2\pi E^2} = \frac{\Delta\mu - E}{kT} \tag{2}$$

After extracting the QFLS from the fit, the entire absorptance spectrum, including the low-energy region, can be reconstructed.[112]

$$\frac{\Phi_{PL}(E)}{\Phi_{bb}(E)exp\,(\frac{\Delta\mu}{kT})} = A(E) \tag{3}$$

The broadening of the onset of the absorptance spectrum, σ, which is related to radiative losses,[28,45,74] is also of interest and is quantified by fitting the first derivative of the absorptance spectrum with a Gaussian function.

$$\frac{dA(E)}{dE} = C_0 + \frac{C_1}{\sigma_{(E_g)}\sqrt{2\pi}}.exp\,(-\frac{(E-E_g)^2}{2\sigma_{(E_g)}{}^2}) \tag{4}$$

Using Lambert-Beer's law, the absorption coefficient can be extracted from A(E), knowing the thickness of the absorber. By linearly fitting the low energy part of the logarithm of the absorption coefficient, the Urbach energy ($E_u$) can be determined (where $\alpha_0$ and $E_U$ are fit parameters). Urbach energies are extracted only for the samples with Na (Mo and HTL) in Figure 5, because for these samples the previously mentioned defect at lower energies is not present in the PL spectra.

$$\alpha(E) = \alpha_o \exp\left(\frac{E}{E_U}\right) \tag{5}$$

From the PL measurements, we extracted the PL quantum yield:

$$Y_{PL} = \frac{\int_0^\infty \Phi_{\mathrm{PL}}(E)\, dE}{F_{Gen}^{Sun}} \approx \frac{\int_0^\infty \Phi_{\mathrm{PL}}(E)\, dE}{F_{Gen}^{Laser}} \tag{6}$$

This is useful to calculate the non-radiative losses[2,26,113] of the absorbers/CdS stacks (Figure S6)

$$\delta\Delta\mu^{nr} = -k_B T ln(Y_{PL}) \tag{7}$$

and the radiative limit (Figure S7).

$$\delta\Delta\mu^{nr} + \Delta\mu = \Delta\mu^{rad} \tag{8}$$

If band-to-band recombination was the only active recombination mechanism and no competing channels were present, the PL intensity would increase linearly with generation flux. In general a power law is observed, where the exponent describes the ODF. A linear fit in log-log plot allows to extract the ODF $\boldsymbol{n_{PL}}$, which is the lower limit of the EDF.[92]

$$G^{n_{PL}} \sim Y_{PL} \tag{9}$$

### 4.3 *J-V*

*J-V* characteristics were measured using a class-AAA AM 1.5 solar simulator at a temperature of 298K (not controlled), with a four-probe configuration. The areas were obtained from an optical microscope. The one diode fits, based on the Orthogonal Distance Regression (ODR)[114], were performed by a home-made script on light J-V characteristics, shifted up by $J_{SC}$.

### 4.4 External quantum efficiency

A homebuilt setup was used to determine EQE spectra of the cells with a four-probe configuration. This contains a lock-in amplifier to measure the photocurrent of the cells under a chopped illumination from a halogen or a xenon lamp passing through a monochromator. The lamp spectrum was determined by Si and InGaAs detectors, with calibrated EQE.

### 4.5 Cathodoluminescence

The CL measurements were conducted at room temperature using an Attolight Allalin 4027 dedicated CL-SEM, equipped with an Oxford Instruments iDus InGaAs 1.7 detector. All CL maps were acquired with an electron beam acceleration voltage of 5 kV, a probe current of 10 nA, and a 100-µm aperture.

### 4.6 Scanning electron microscopy

The cross-sectional microstructures of the cut films were analysed by different scanning electron microscopes. Cross-section images were taken with an acceleration voltage of 7 kV with a SEM Hitachi SU-70. The compositions were obtained by energy-dispersive X-ray spectroscopy (EDS) in a Zeiss EVO-10 on the bare absorbers at an acceleration voltage of 20kV and on the HTLs at 10kV.

### 4.7 Raman spectroscopy

Raman spectroscopy was performed with a Renishaw inVia micro-Raman spectrometer equipped with a 633 nm excitation laser source and a 2400 lines $mm^{-1}$ grating.

### 4.8 Grazing incidence X-ray diffraction

GIXRD patterns were acquired with an incidence angle a of 0.2° on a Bruker D8 Discover diffractometer (Bruker) using Cu-Kα radiation in the 2Θ range from 20° to 50°.

### 4.9 Transmission electron microscopy

Cross-section TEM samples were prepared using a focused ion beam FEI Helios Nanolab 650 and analysed with a JEOL F200-Cold FEG. Elemental mapping and line scan were obtained by EDS in STEM mode.

### 5.0 Confocal microscope

The surface was analysed using a Sensofar S Neox 090 3D optical profiler, which utilizes non-contact measurement technique. Roughness was quantified in accordance with the ISO 25178 international standard. To isolate roughness from other surface components, a standardized filtering process with three filters was applied in compliance with ISO 25178.

### 5.1 Secondary ion mass spectrometry

SIMS measurements were carried out on a CAMECA SC-Ultra instrument. The films were analysed using a $Cs^+$ low-impact energy beam (1 keV) in the depth-profiling mode. The variation of the elemental composition was acquired across the whole thickness of the films, with the intensity of the elements collected from a 60-μm area in diameter centred in a 250 μm × 250 μm scanned area.

**Table 1.** List of symbols.

| Symbol | Meaning (unit) | Symbol | Meaning (unit) |
|---|---|---|---|
| $E$ | Photon Energy (eV) | $\Phi_{BB}(E)$ | Black body emission spectrum ($1 \cdot cm^{-2}s^{-1}V^{-1}$) |
| $E_g$ | Band gap energy (eV) | $J_{SC}$ | Short circuit current density ($A \cdot cm^{-2}$) |
| $F_{Gen}^{Sun}$ | Sun generation Flux ($cm^{-2}s^{-1}$) | $Y_{PL}$ | PL quantum yield |
| $F_{Gen}^{Laser}$ | Laser generation Flux ($cm^{-2}s^{-1}$) | $V_{OC}$ | Open circuit voltage (V) |
| $E_U$ | Urbach energy (eV) | $\Delta\mu$ | Quasi Fermi level splitting (QFLS) (eV) |
| $A(E)$ | Absorptance spectrum from PL | $\Delta\mu^{rad}$ | Radiative limit of QFLS (eV) |
| $k_B$ | Boltzmann constant | $\delta\Delta\mu^{nr}$ | Non-radiative QFLS loss (eV) |
| $T$ | Temperature (K) | $V_{OC}$ | Open-circuit voltage (V) |
| $G$ | Generation flux ($cm^{-2}s^{-1}$) | $\Phi_{PL}(E)$ | Photoluminescence emission spectrum ($1 \cdot cm^{-2}s^{-1}V^{-1}$) |
| $\alpha$ | Absorption coefficient ($cm^{-1}$) | $FF$ | Fill factor |
| $d$ | Sample thickness | $R_s$ | Series resistance ($\Omega \cdot cm^{-2}$) |
| $\sigma$ | Absorptance edge broadening parameter (meV) | $R_{sh}$ | Shunt resistance ($\Omega \cdot cm^{-2}$) |
| $n_{PL}$ | Optical diode factor (ODF) | $J_0$ | Saturation current density ($A \cdot cm^{-2}$) |

**ASSOCIATED CONTENT**

The Supporting Information is available free of charge.

**Data availability**

All data supporting this study are available in Zenodo.

**Notes**

The authors declare no competing financial interest.

**Author Contributions**

F.L. (lead) designed the experiments, prepared the CISe absorbers, measured and analysed the samples by PL, conducted and analysed the *J-V* and EQE measurements, made the simulation and wrote the manuscript. Z.Z. conducted the *J-V* and EQE measurements, prepared the HTLs, optimized their sputter deposition and wrote the part discussing it. B.K. and S.B. performed

SEM and XRD measurements. A.M. and H.Y. contributed with confocal microscope and Raman measurements. Y.H., G.K. and R.O. measured and analysed samples by CL and SEM. N.V. and A.P. conducted the TEM and SIMS measurements and analysis. M.M contributed to the most parts of the devices including glass cleaning, preparation of Mo, CdS, TCO and grids deposition. M.M also conducted some of the SEM measurements. S.S. defined and initiated the project, acquired the funding, supervised the work, guided data acquisition and modelling and contributed to the writing of the paper. All authors contributed to results discussion and to revise the paper.

**ACKNOWLEDGMENTS**

Financial support is acknowledged from the Luxembourgish Fonds National de la Recherche (FNR) in the framework of the project FULL CORE/22/MS/17138895/FULL, from UKRI by the REACH project INTER/UKRI/20/15050982, from the European Union in the framework of the HiBITS project and from ERA-NET in the framework of the HIDDEN-PV project. 

**What enables $GaO_x$ as hole transport layer for a 16% 1.0 eV $CuInSe_2$ Bottom Cells with $V_{OC}$ above 550 mV?**

*Francesco Lodola*, Zhuangyi Zhou, Boaz Koren, Saeed Bayat, Alessandro Magon, Yucheng Hu, Adrian-Marie Philippe, Michele Melchiorre, Hasan Arif Yetkin, Gunnar Kusch, Nathalie Valle, Rachel A. Oliver and Susanne Siebentritt**

Francesco Lodola, Zhuangyi Zhou, Boaz Koren, Saeed Bayat, Alessandro Magon, Michele Melchiorre, Hasan Arif Yetkin, Susanne Siebentritt
University of Luxembourg, Department of Physics and Materials Science, 41 Rue du Brill, Belvaux, L-4422 Luxembourg

Yucheng Hu, Gunnar Kusch, Rachel A. Oliver
Department of Materials Science and Metallurgy, University of Cambridge, 27 Charles Babbage Road, Cambridge CB3 0FS, United Kingdom

Adrian-Marie Philippe, Nathalie Valle
Luxembourg Institute of Science and Technology (LIST), 41, rue du Brill, L-4422, Belvaux, Luxembourg

*E-mail: francesco.lodola@uni.lu, susanne.siebentritt@uni.lu

Supplementary information.

**Figure S1**. Cathodoluminescence comparison between the sample CIGSe and CISe HTL. a) panchromatic cross-section CL intensity map (colour code in arbitrary units); b) centre wavelength of the fitted near band edge CL emission from the cross section; c) panchromatic plan view CL intensity map (colour code in arbitrary units); d) centre wavelength of the fitted near band edge CL emission from the plan view. A gradient in the energy of the CL emission is clearly visible in the CIGSe sample in the cross-section emission energy map (b).

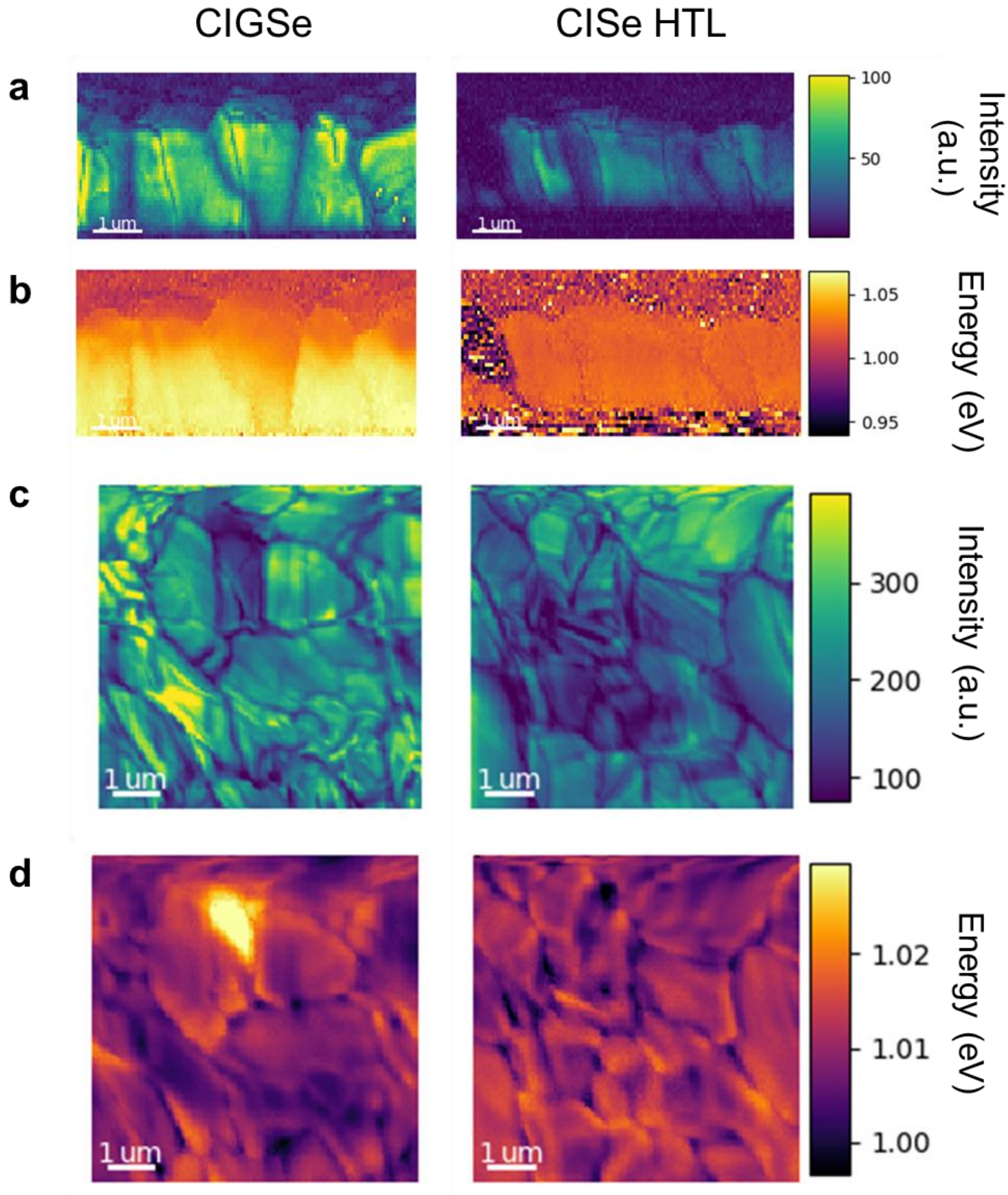

**Figure S2**. SEM cross-section images with different fields of view. Samples grown using the process "LT" still exhibit the two layers of HTL on top of the columnar molybdenum grains, after the absorber deposition.

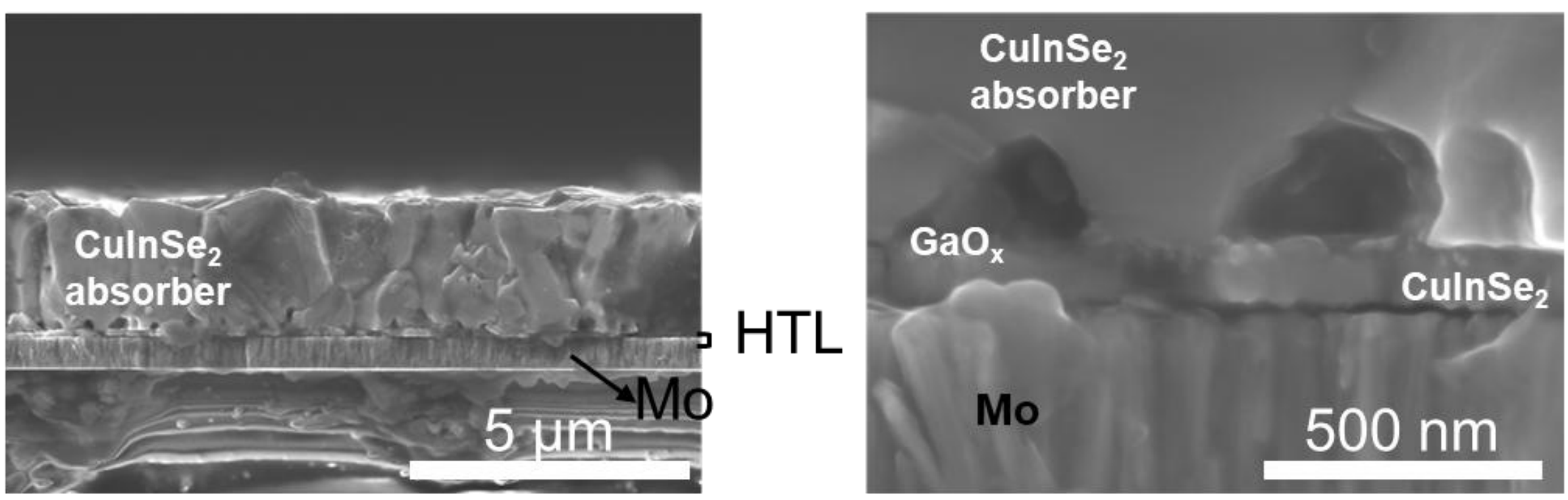

**Figure S3**. Confocal microscope top view with roughness values (average "Sa" and root mean square "Sq" values) for a) SLG/Mo/HTL/CISe stack grown with "Cu HT" and; b) SLG/Mo/HTL/CISe stack grown with "LT". Statistic of Sa and Sq roughness for c) HTLs ("LT" and "Cu HT") compared to Mo samples; d) HTL samples "LT" versus "Cu HT".

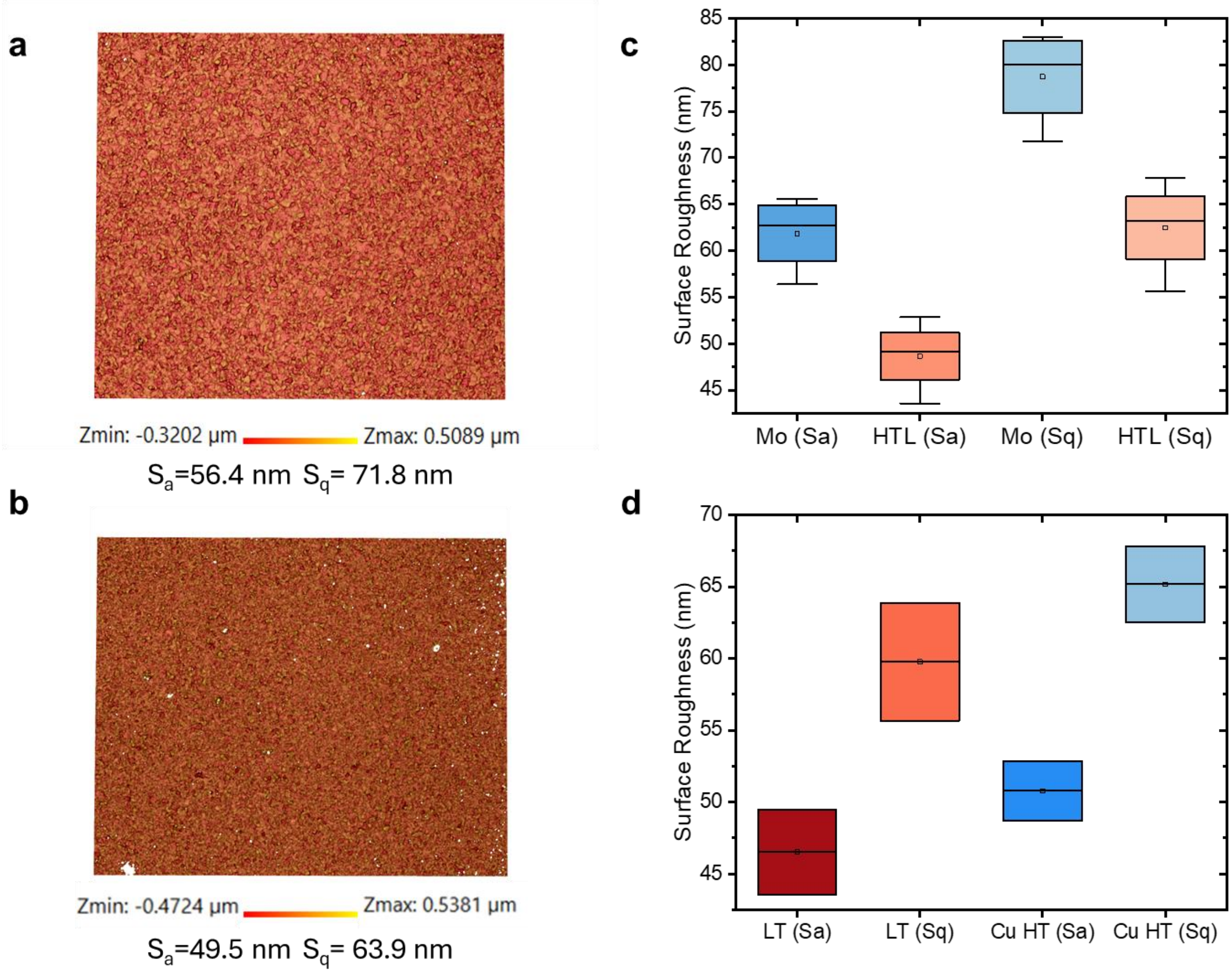

**Figure S4**. SEM-Energy-dispersive X-ray spectroscopy (EDS) spectrum of the two layers, composing the HTL, probed by an electron beam with a landing energy of 10 keV through the oxide layer. The atomic percentages of the elements are reported in the table. The result shows that the $CuGaSe_2$ layer was Cu-poor.

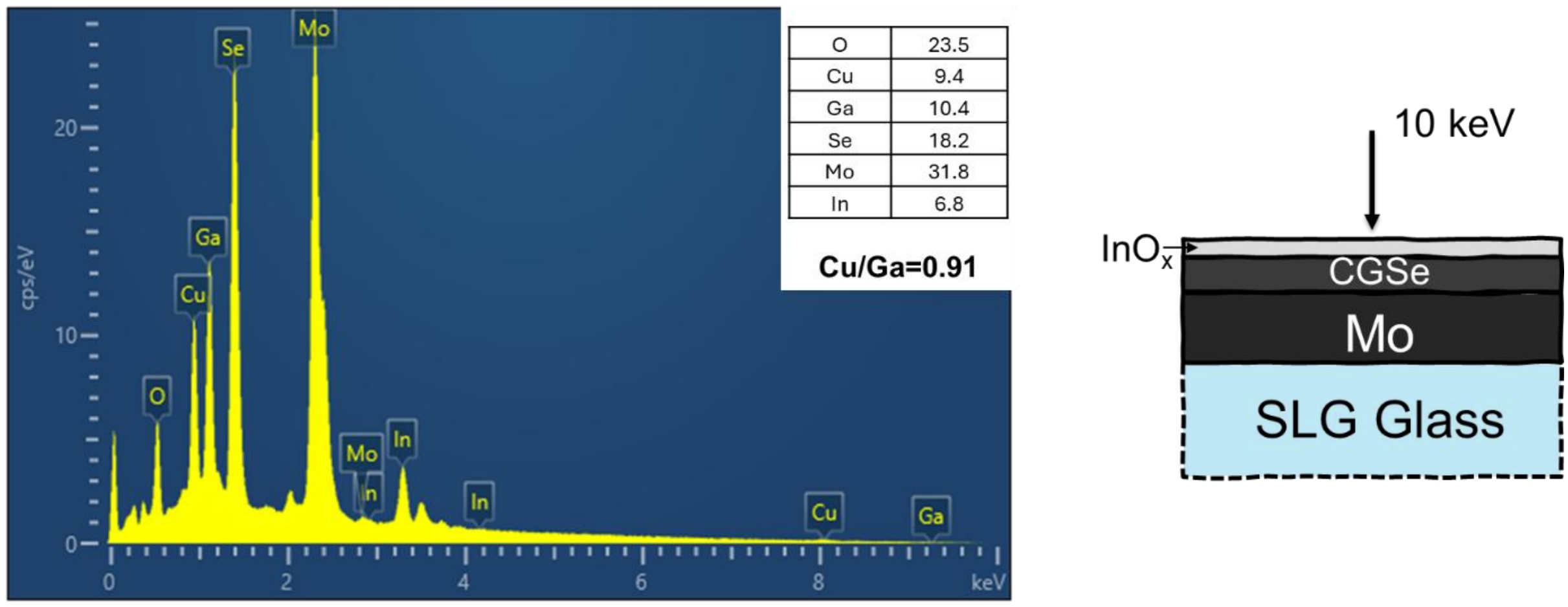

**Figure S5**. The element distribution along the cross-section near the backside of the HTL "LT" sample (without Na barrier) giving the record device, measured by the EDS-X line scanning. This confirms the In and Ga ion-exchange between the two layers of the HTL.

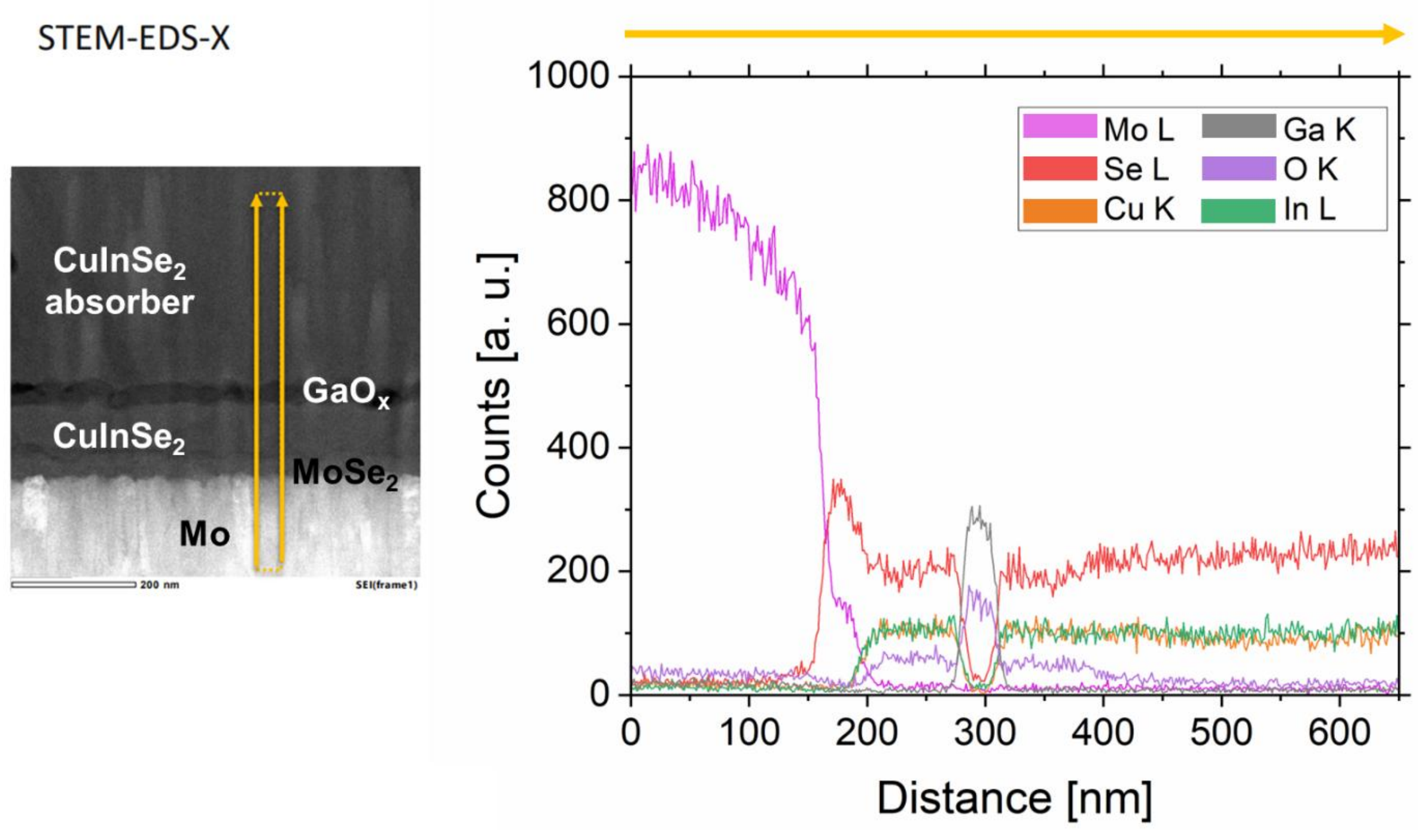

**Figure S6**. a) non-radiative losses (from Equation 7); b) fill factors; c) σ, the absorptance broadening (from Equation 4) as a function of Cu/In of the various absorbers.

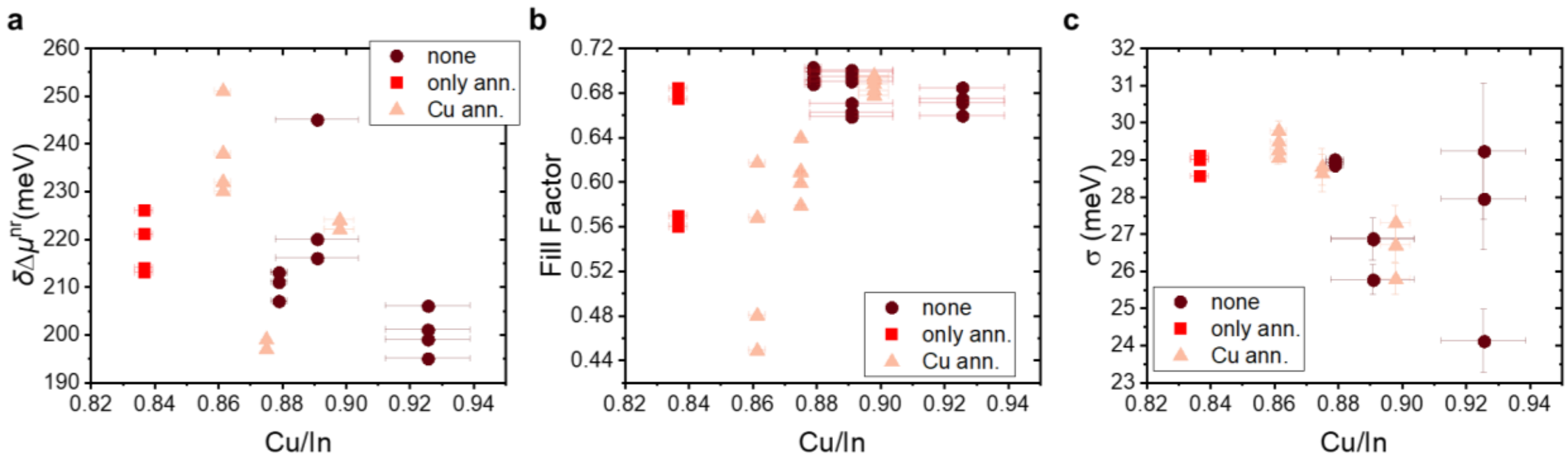

Non-radiative losses are influenced by the Cu content of the absorbers. For overly Cu-poor compositions, there is a significant increase in voltage loss, more pronounced in the "Cu HT" samples. Fill factors are less affected by variations in Cu/In for the "HT" and "LT" samples, while "Cu HT" shows a sharp drop when Cu/In falls below 0.88. The absorptance edge broadening does not show any trend with Cu/In for the "LT" and "HT" samples, although only low values of σ are observed for Cu/In above 0.88. In contrast, for "Cu HT", both values decrease as the absorber approaches stoichiometry.

**Figure S7**. Bandgap energy ($E_g$), extracted from the Gaussian fit of the first derivative of A(E) (Equation 4), minus the radiative limit ($\Delta\mu^{rad}$) from Equation 8 of the absorbers grown with the three different processes as a function of σ the absorptance broadening. The radiative limit is defined as the sum of the QFLS, obtained from the fitting of the PL spectra, and the non-radiative losses, extracted from the photoluminescence quantum yield. Using Eg for this comparison instead of the corresponding Shockley-Queisser $V_{OC}$ is justified, because the bandgaps change only by a few meV.

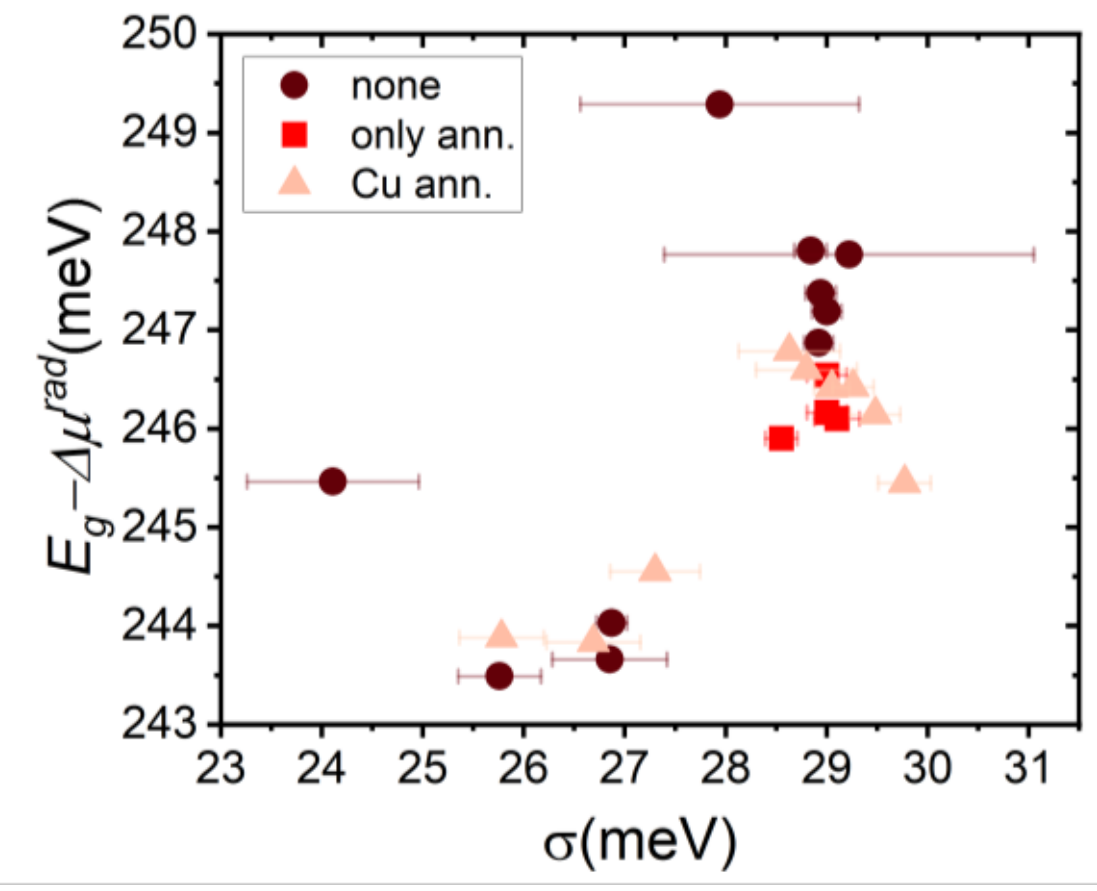

**Figure S8**. Short-circuit density for "LT", "HT" and "Cu HT" devices. No differences are noted.

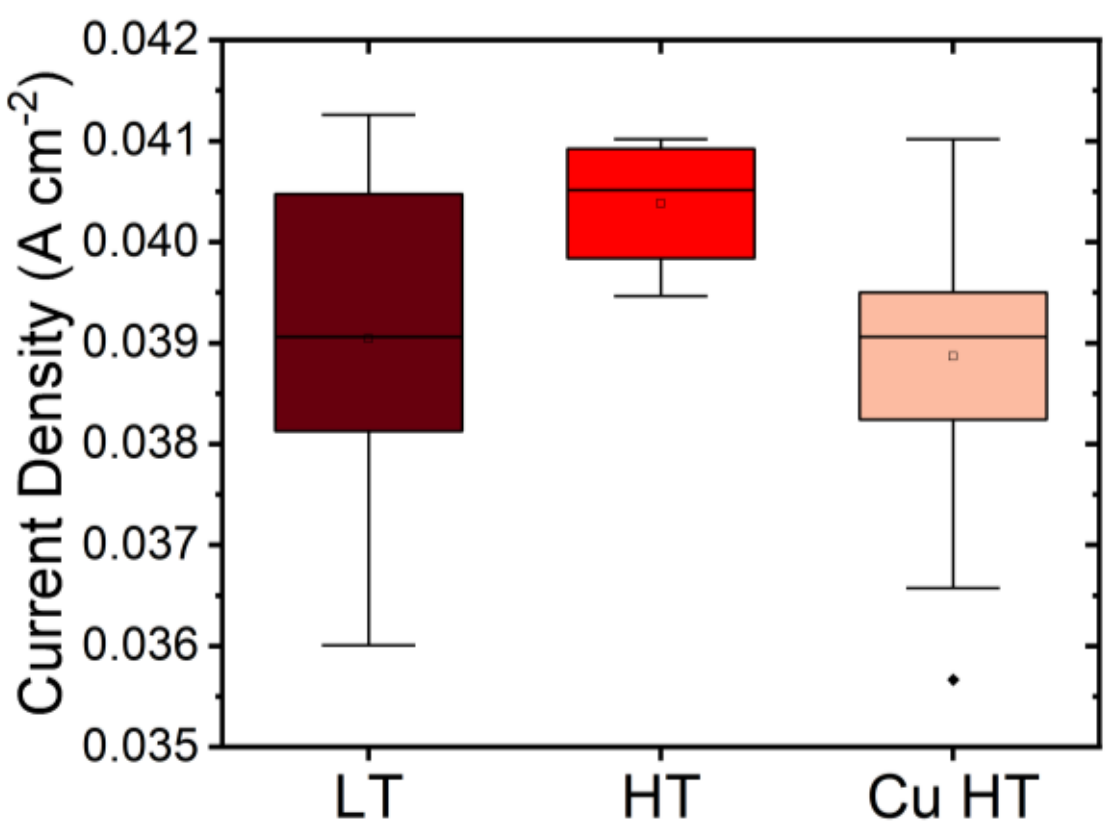

**Figure S9**. The effect of the sodium precursor, deposited before or after the HTL, on the absorbers' PL. Additional sodium is detrimental for the absorber. This suggests that sodium can, at least, partially diffuse through the oxide layer. The quantity of Na in the absorber from the soda-lime glass substrate is sufficient, in contrast to what was formerly reported[1], likely due to the use of higher-Na-content glass.

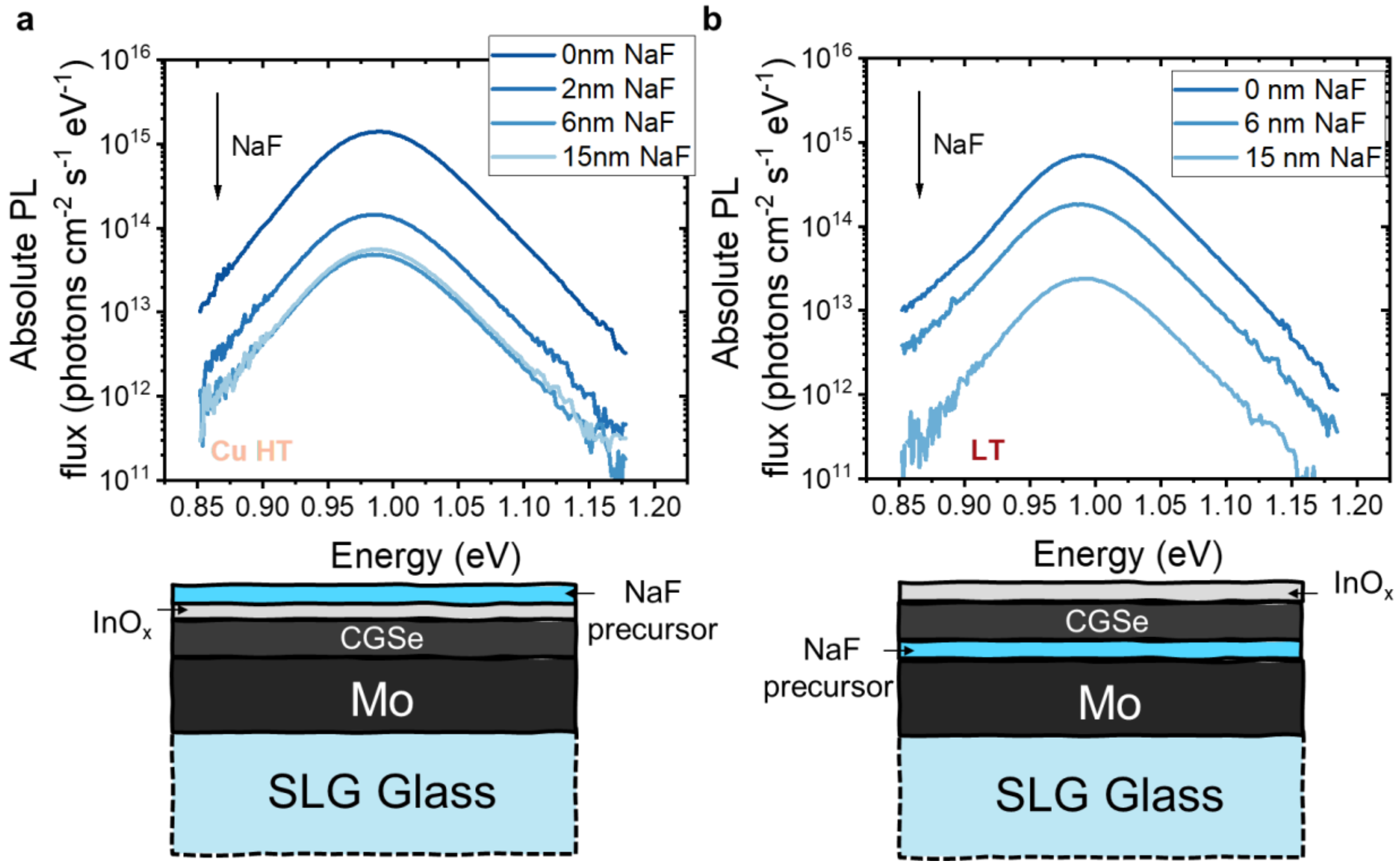

**Figure S10**. SIMS measurements of “LT” HTL passivated absorbers with a) and without b) Na barrier. Normalized intensities plots are displayed. The film without a Na barrier is In-rich at its surface.

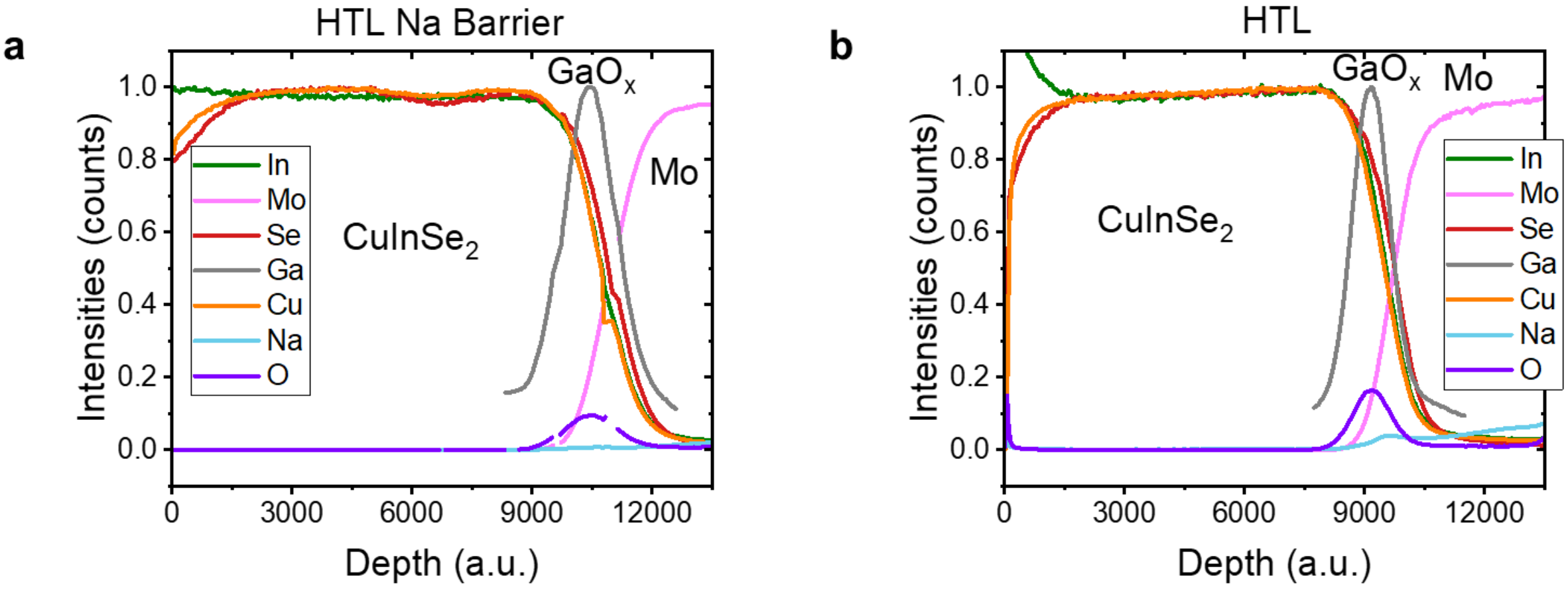

SIMS measurements confirm the $GaO_x$ formation on HTL, with and without Na barrier by the clear Ga peak located at the absorber backside. In the HTL Na Barrier sample the presence of O (with Si, not displayed here for clarity reasons) in a layer between the soda-lime glass (right) and Mo back contact, verifies the sputtered $SiO_x$ blocking layer. Na gradient is notable in Mo.

**Figure S11**. Comparison of Na SIMS profiles with (black) and without Na barrier (red).

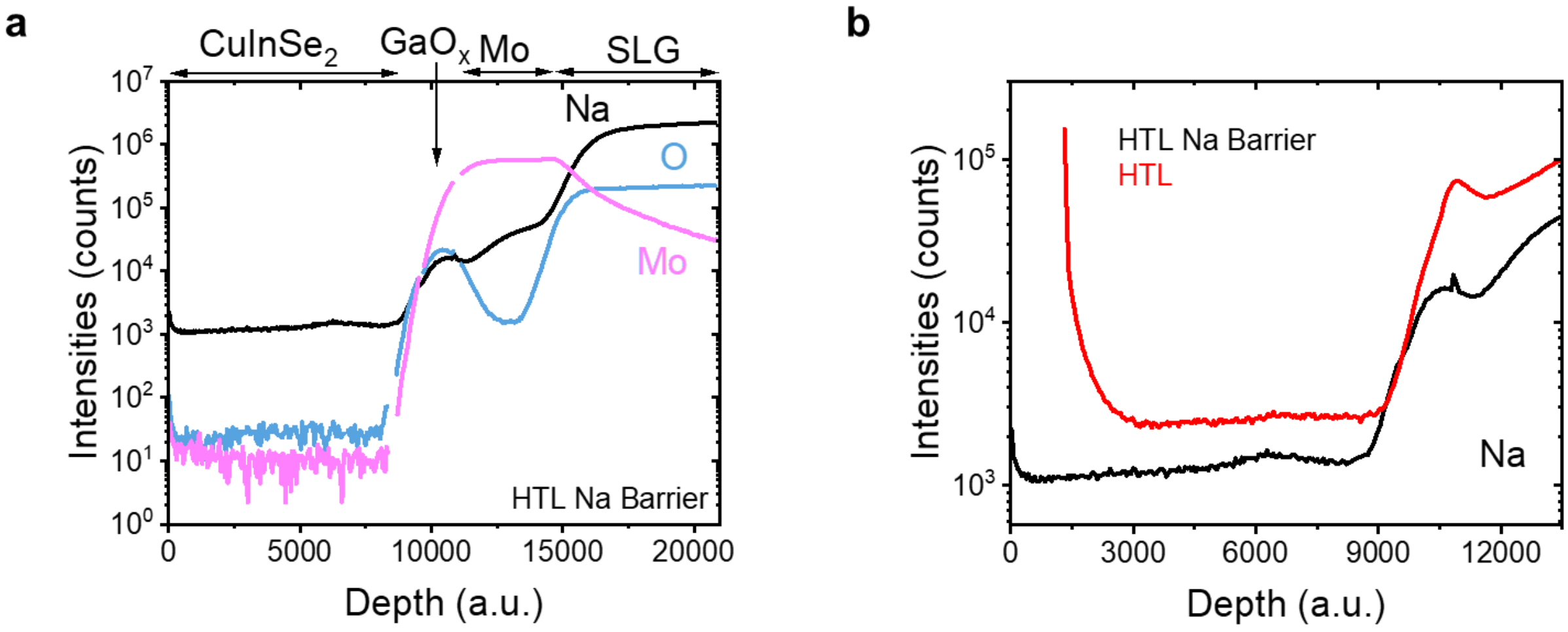

Na differences are notable between the samples with and without Na barrier. However, the HTL Na barrier sample still contains Na. Smaller grains in the sample with Na barrier are the considered the main cause, as explained in the main text. Contamination from the growth chamber and from $Na_2Se$ volatile species from the SLG substrates during the growth are thought to be another cause.[2] The $GaO_x$ layer shows a partially Na blocking behaviour.

**Figure S12**. SEM cross section images of HTL without (left) and with (right) Na blocking barrier, with different fields of view. Both exhibit the two layers of HTL on top of molybdenum, after the absorber deposition. The HTL sample with Na barrier shows smaller grains than the one grown without barrier.

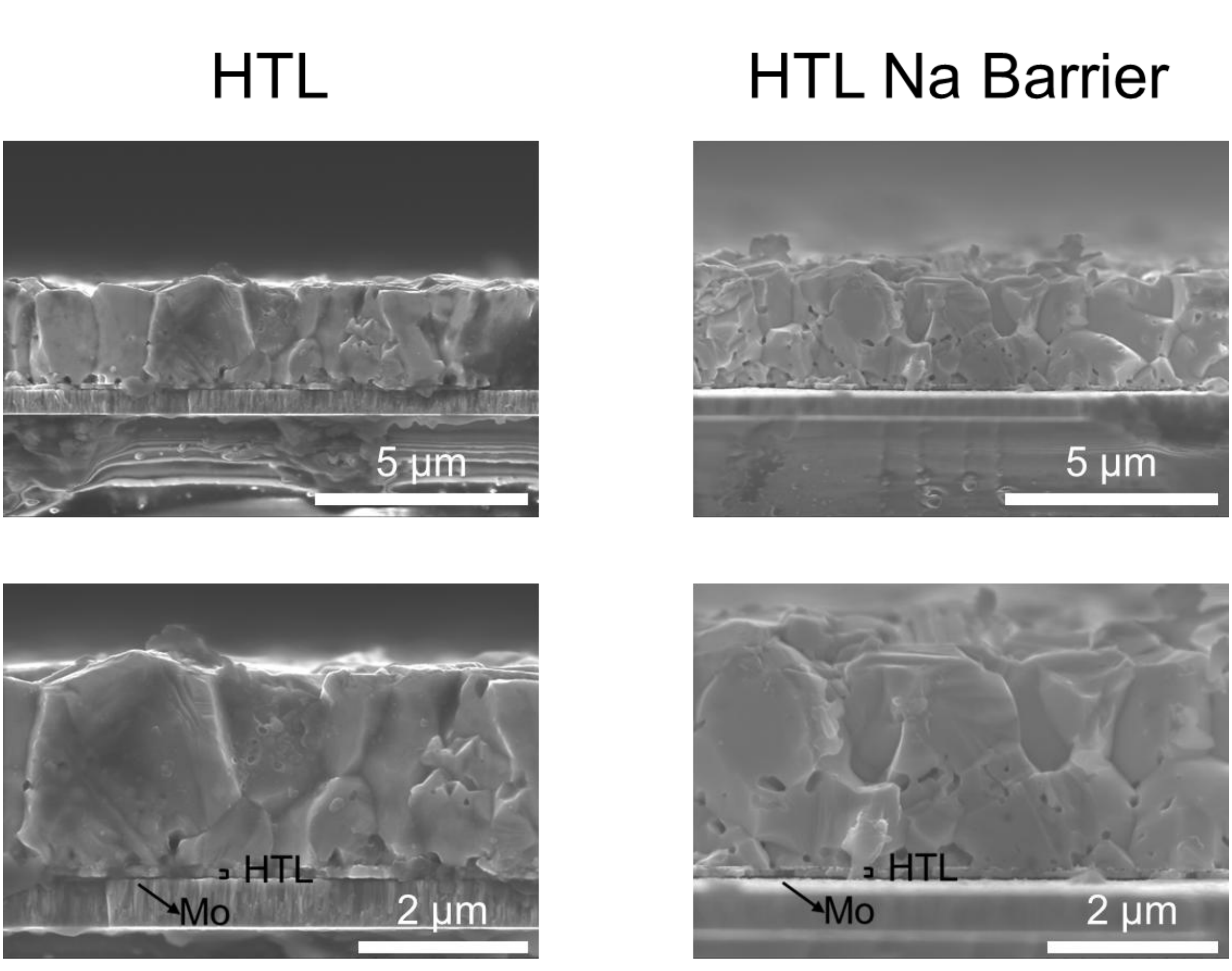

**Figure S13.** Absorption coefficient of samples in Figure 5, with the values of the Urbach energies ($E_u$) (Equation 5). The absence of Na leads to presence a defect low energy. Therefore, $E_u$ can be evaluated only for the samples without Na barrier. Here we can see that the passivation of the back contact reduces this sub-bandgap states. This might be due to lower recombination at the back side, at the interface with the metallic back contact, which is particularly defective and contributes to an increase in static disorder. Fittings are performed in the range where the absorption coefficient is below 10 $cm^{-1}$.[4] For the sample HTL the lower fitting limit is 0.87, to avoid additional noise where absorption coefficient is around $10^{-1}$ $cm^{-1}$.

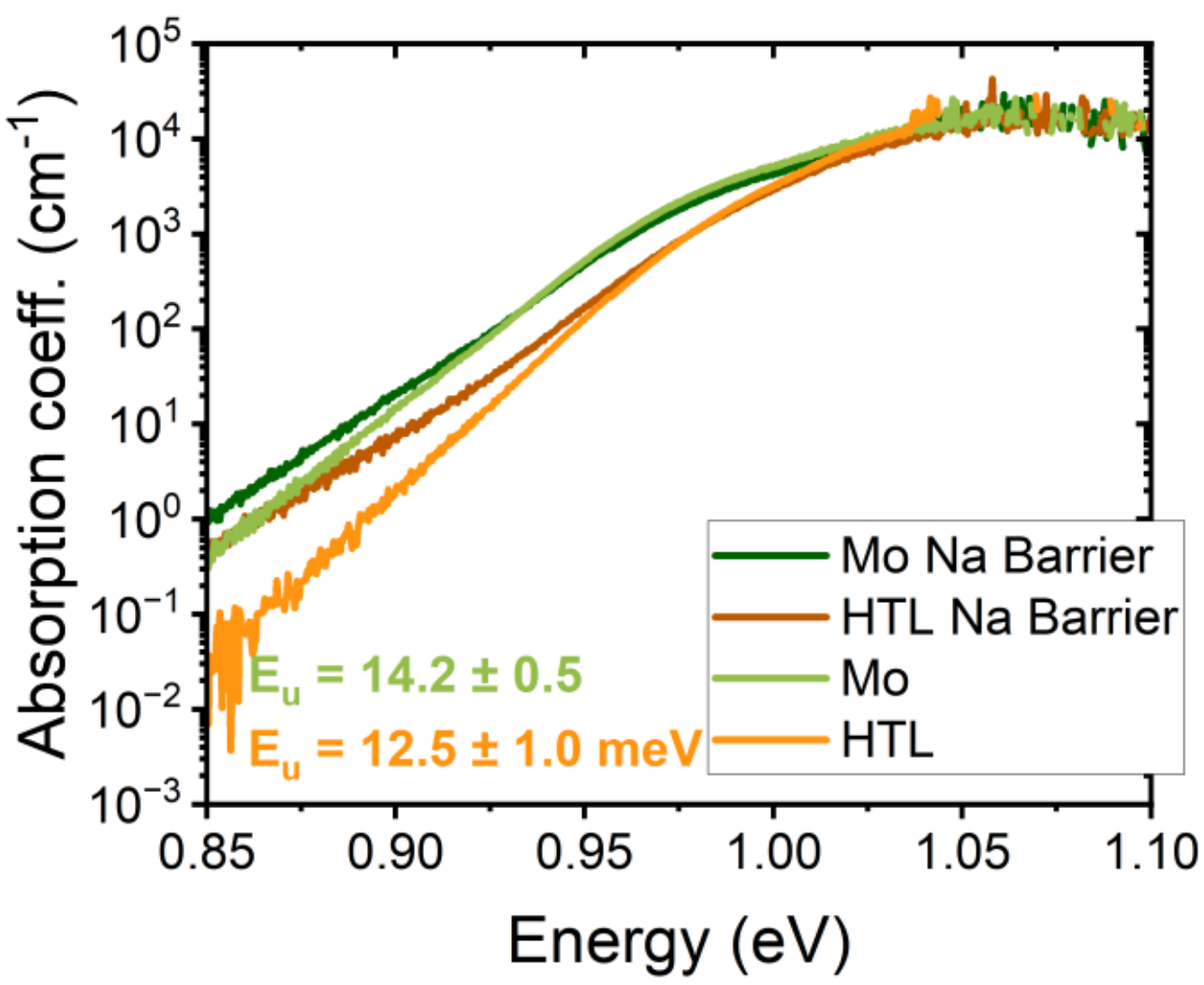

**Figure S14**. CISe HTL with Na barrier: a) dark-field STEM cross-sectional image near the backside (Mo at the bottom) and; b) corresponding elemental mapping measured by EDS showing the distribution of Cu, In, Se, Mo, O and Ga. Ion-exchange is also present in absence of sodium.

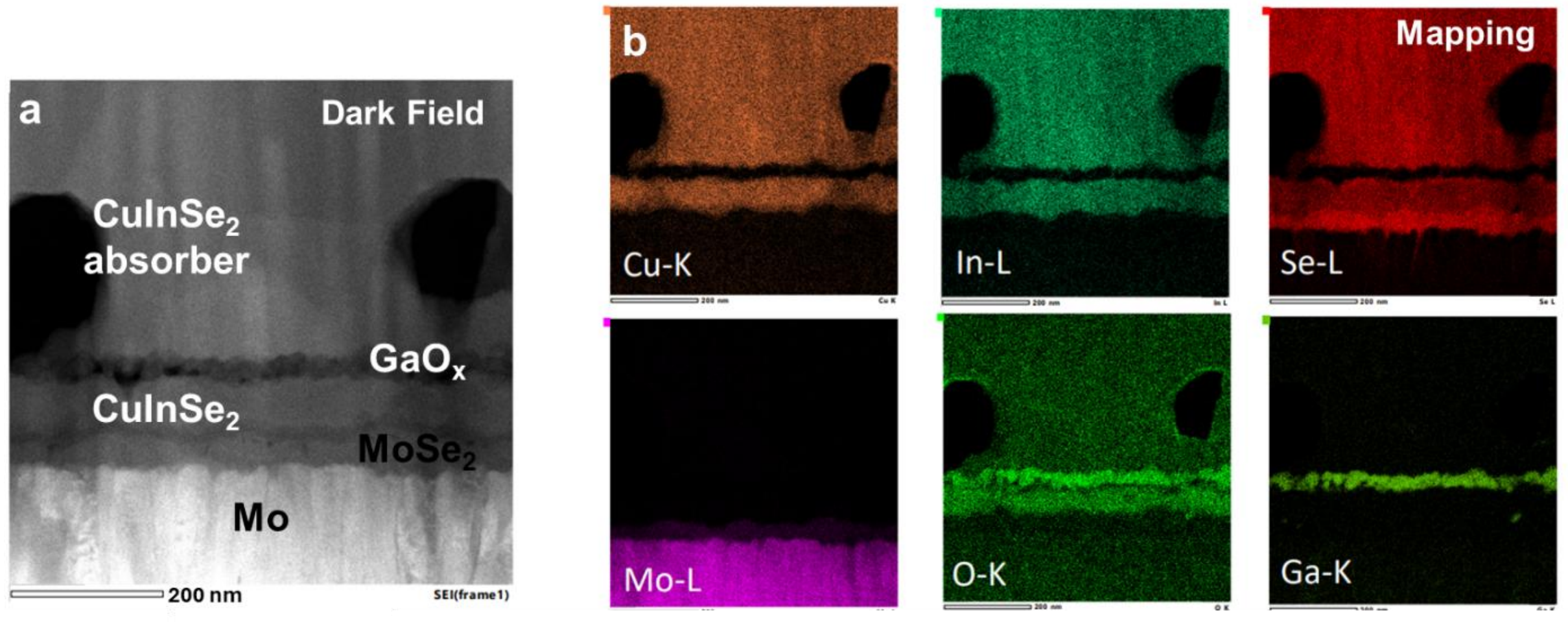

**Figure S15.** a) PL absorptance (Equation 3) of samples in Figure 5; b) first derivative of EQE of the same samples (Figure 5b), fitted with a Gaussian function, fitting range in grey (Equation 4).

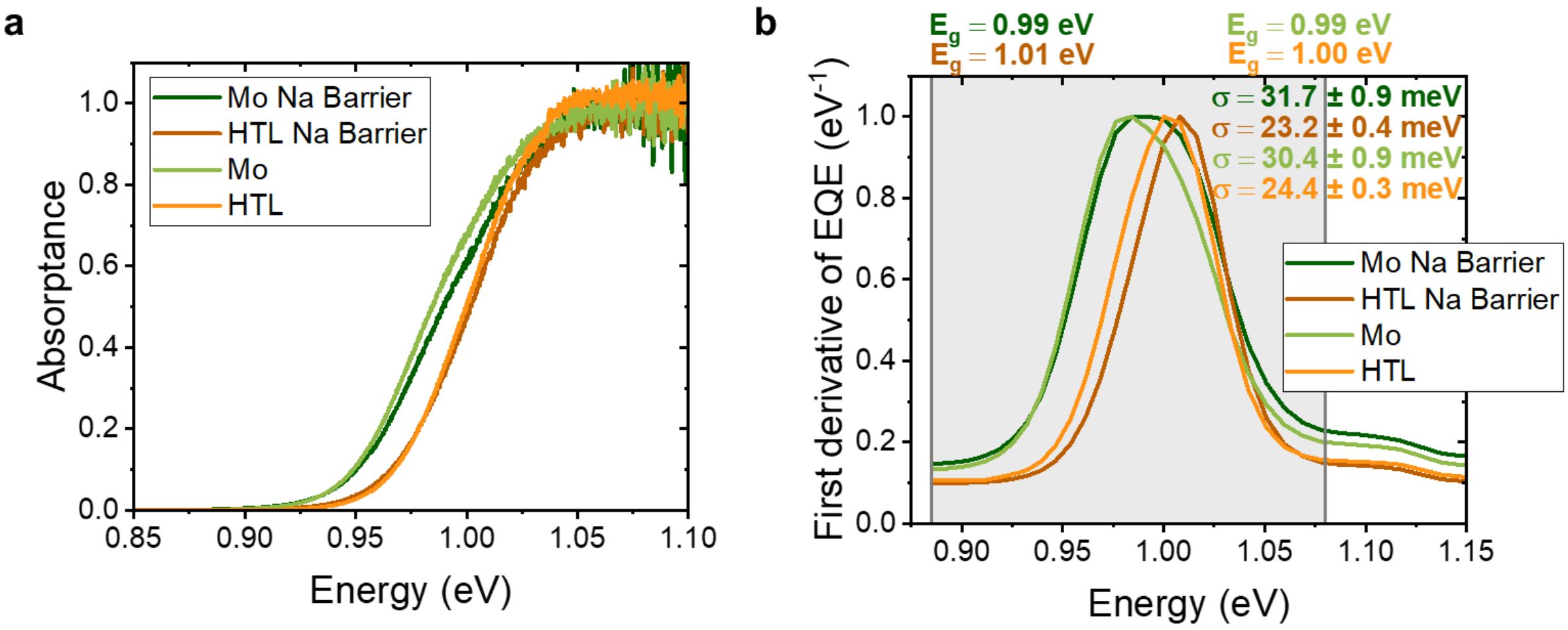

**Figure S16**. SEM-Energy-dispersive X-ray spectroscopy spectrum of the absorber layers for Na experiment. All samples are Cu-poor.

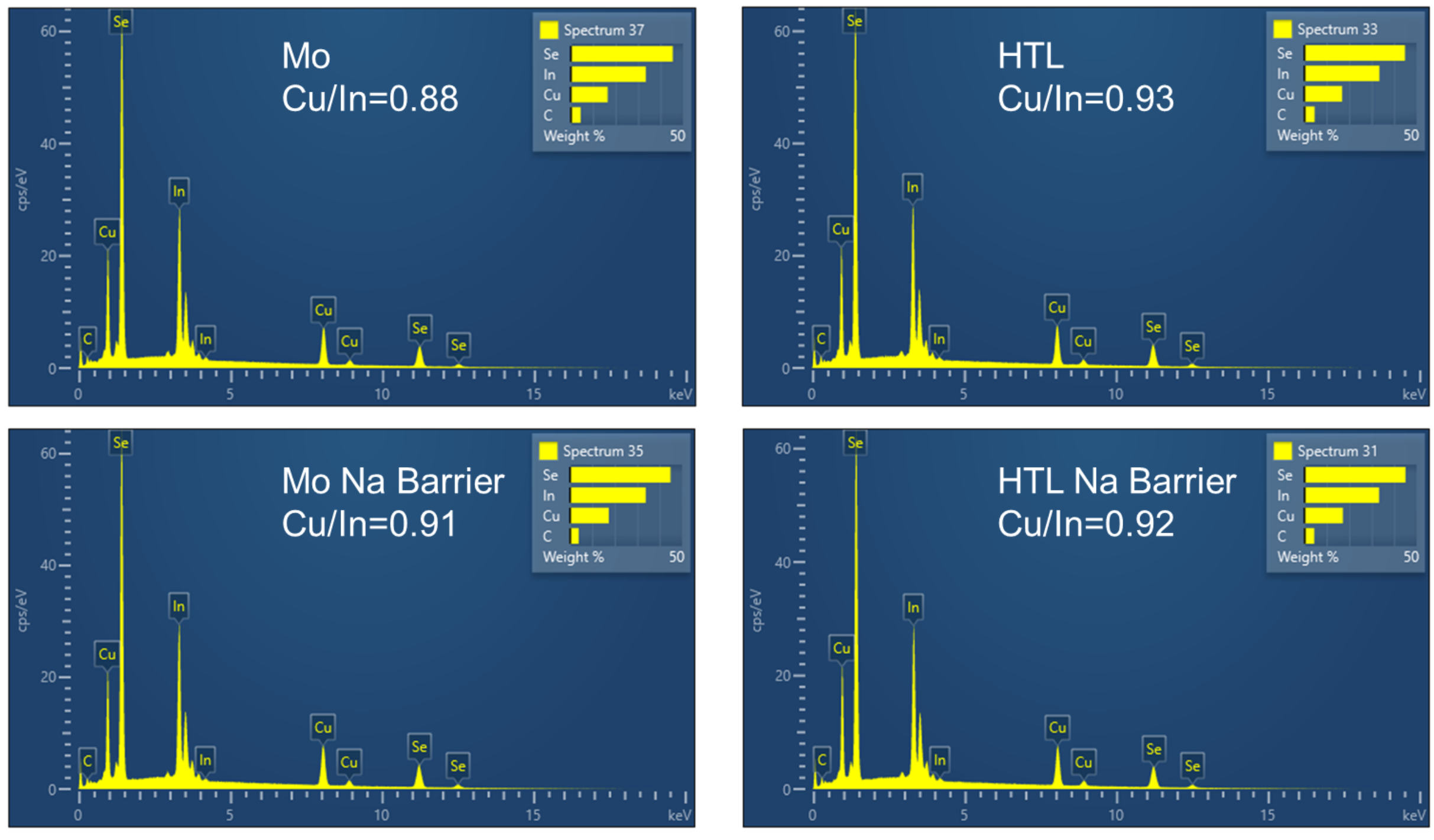

**Figure S17**. HTL sample with Na barrier: High-resolution TEM images of the back-side region with Fast Fourier transform (FFT) of an area (green square region) in the $GaO_x$. The oxide

layer is also partially crystalline in absence of sodium from SLG substrate. The layers composing the HTL maintain the same structure of sample with Na (Figure 6).

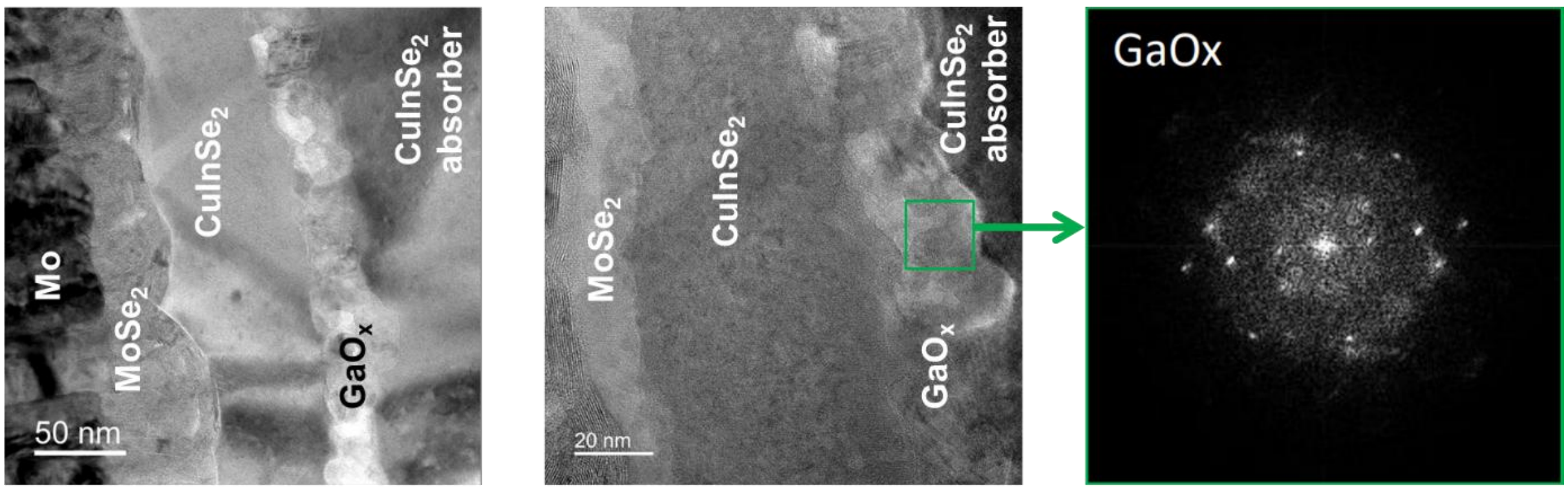

**Figure S18**. Grazing Incidence X-ray Diffraction (GI-XRD) at angle of 0.2° of sputtered polycrystalline 100 nm $InO_x$ (dark blue) and 30 nm $InO_x$ (light blue) on glass substrate, the HTL structure (grey) and the HTL structure annealed at 500 C° for 24 minutes in presence of 10 nm NaF(red). At the bottom, the reference lines for the various compounds: $CuInSe_2$[5], $CuGaSe_2$[6], $Ga_2O_3$[7], $In_2O_3$[8].

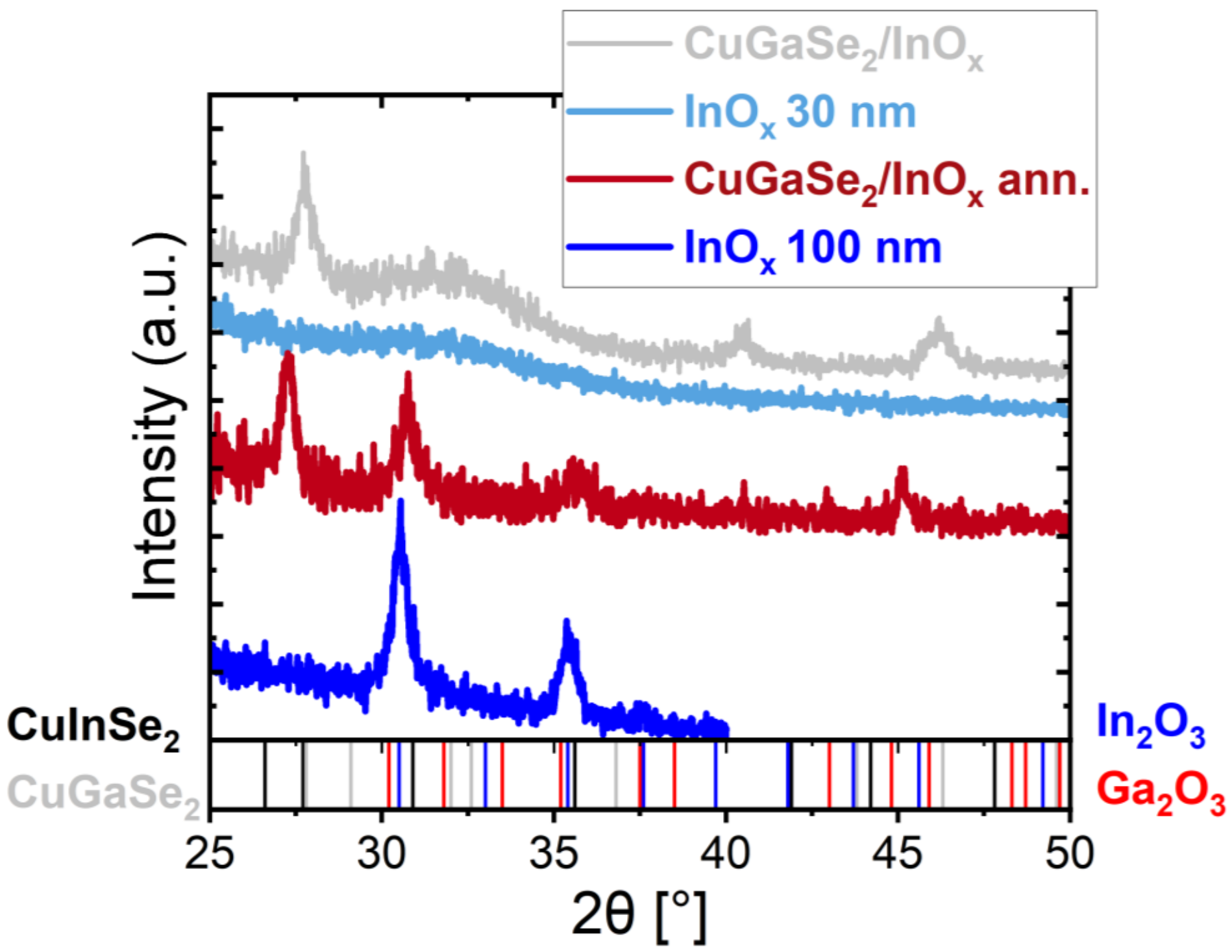

It is interesting to note that the initially sputtered 30 nm $InO_x$ on $CuGaSe_2$ is amorphous (no InOx peak visible in light grey diffractogram). This is compatible with XRD diffractograms of the same sputtered 30 nm $InO_x$ on glass substrate. Only thicker films of sputtered $InO_x$ have crystalline properties (dark blue). After annealing, the sputtered $InO_x$ becomes partially crystalline, with new peaks appearing, and the ion-exchange between the $CuGaSe_2$ and the oxide layer starts to happen (see the shift belonging to the $CuGaSe_2$ layer moving towards lower angle).

**Figure S19**. Raman spectroscopy, probed with a 633 nm laser through the $InO_x$ layer, of the HTLs: as-deposited (blue), annealed in vacuum at 500 °C for 24 minutes without (bright red) and with additional Na presence (dark red). Only the chalcopyrite phase shows signal. The shift in the $A_1$ mode of chalcopyrite suggests ion-exchange between the two layers of the HTL[9,10], with the chalcopyrite layer incorporating more indium ($CuInSe_2$ absorber is shown as a reference). The ion-exchange is further catalysed by the presence of additional Na, introduced as a 10 nm precursor on top of the oxide layer. It is known that Na can enhance the formation of $GaO_x$.[3] We suggest here that, since the two layers are very thin and there are fewer grain boundaries, Na promotes intragrain In–Ga interdiffusion, as previously reported.[11]

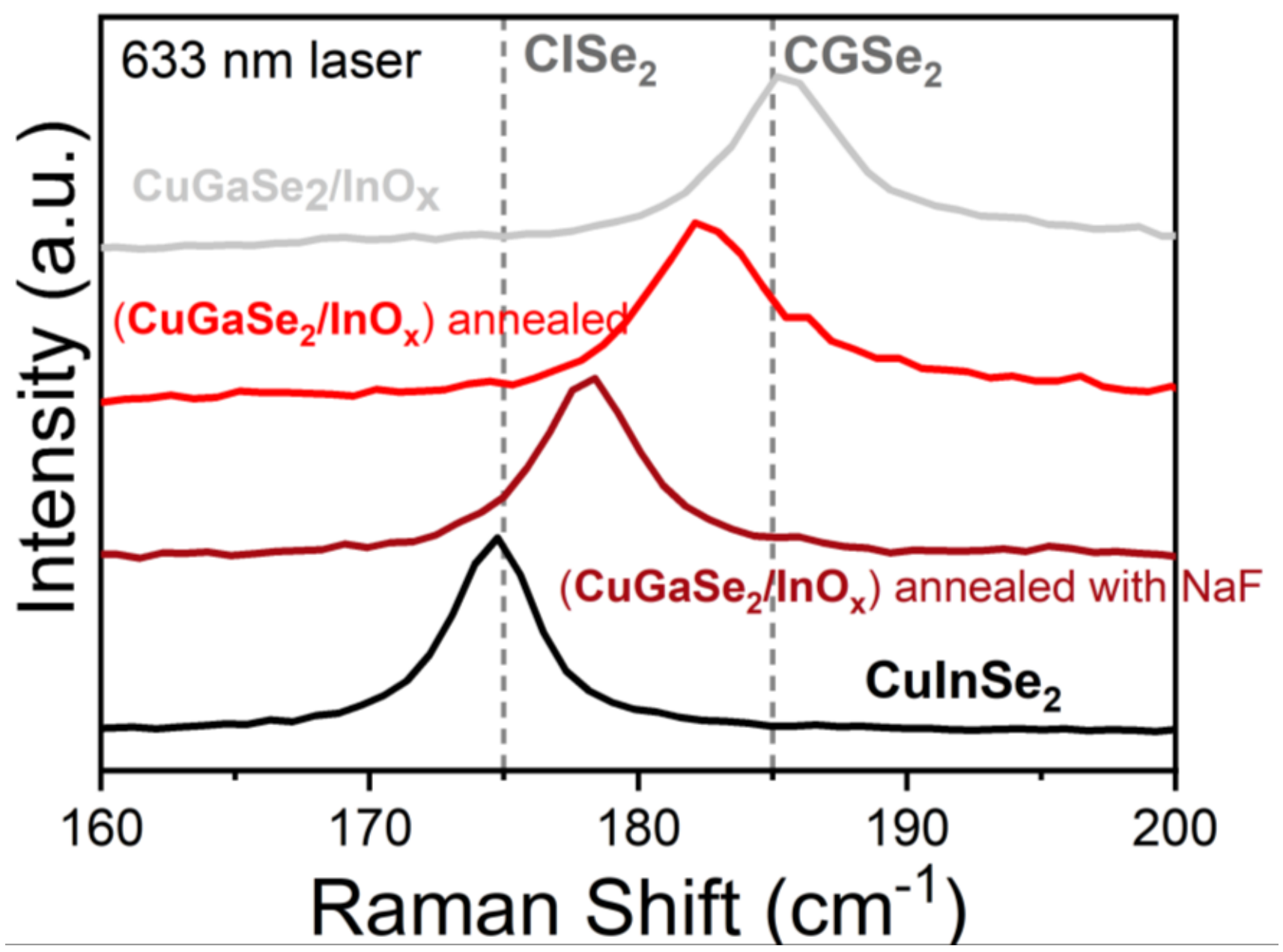

**Figure S20**. Examples of one diode fits on “LT” and “Cu HT” solar cells. Both show high electric diode factor values.

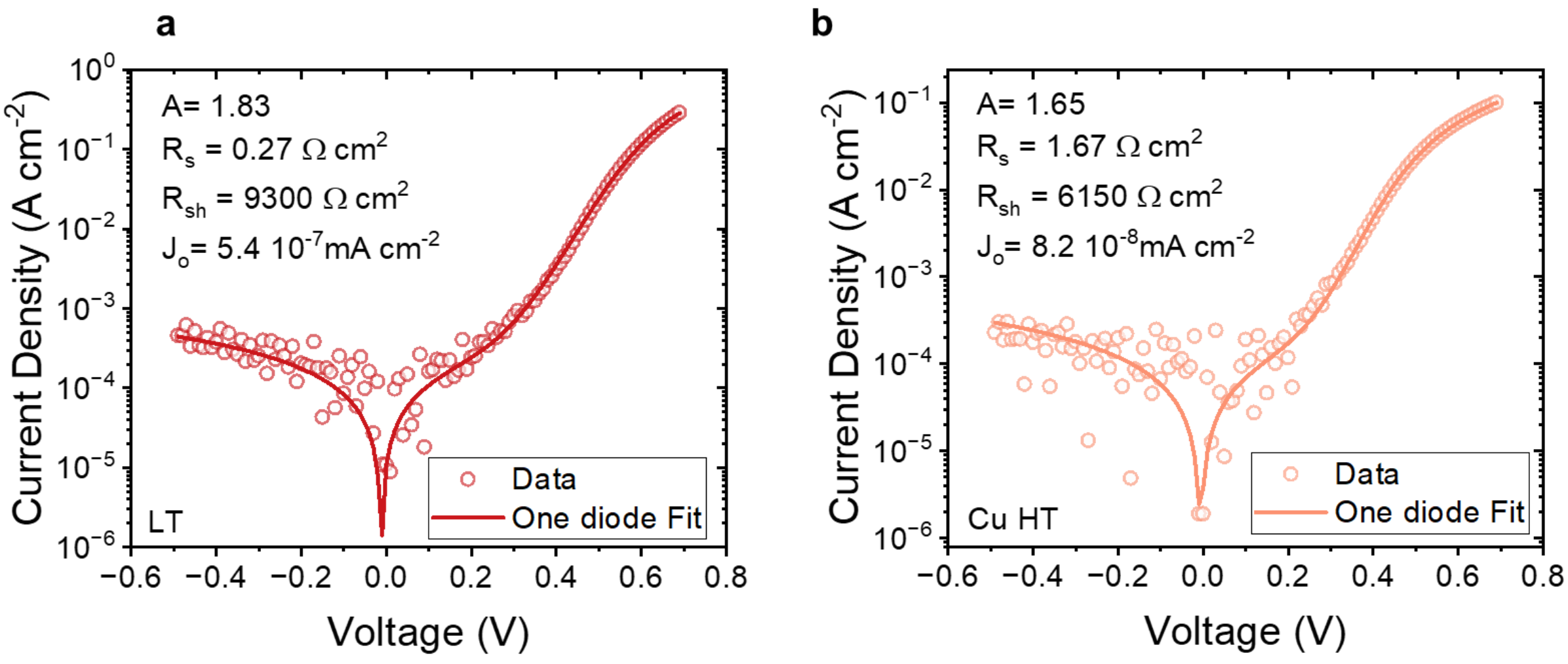

**Figure S21**. Certification report of "LT" $CuInSe_2$ solar cell with 1.00 eV bandgap by Fraunhofer ISE. It has the record-certified $V_{oc}$ of 552 mV for ungraded $CuInSe_2$[12,13], opening up possibilities for this device structure. Device area:

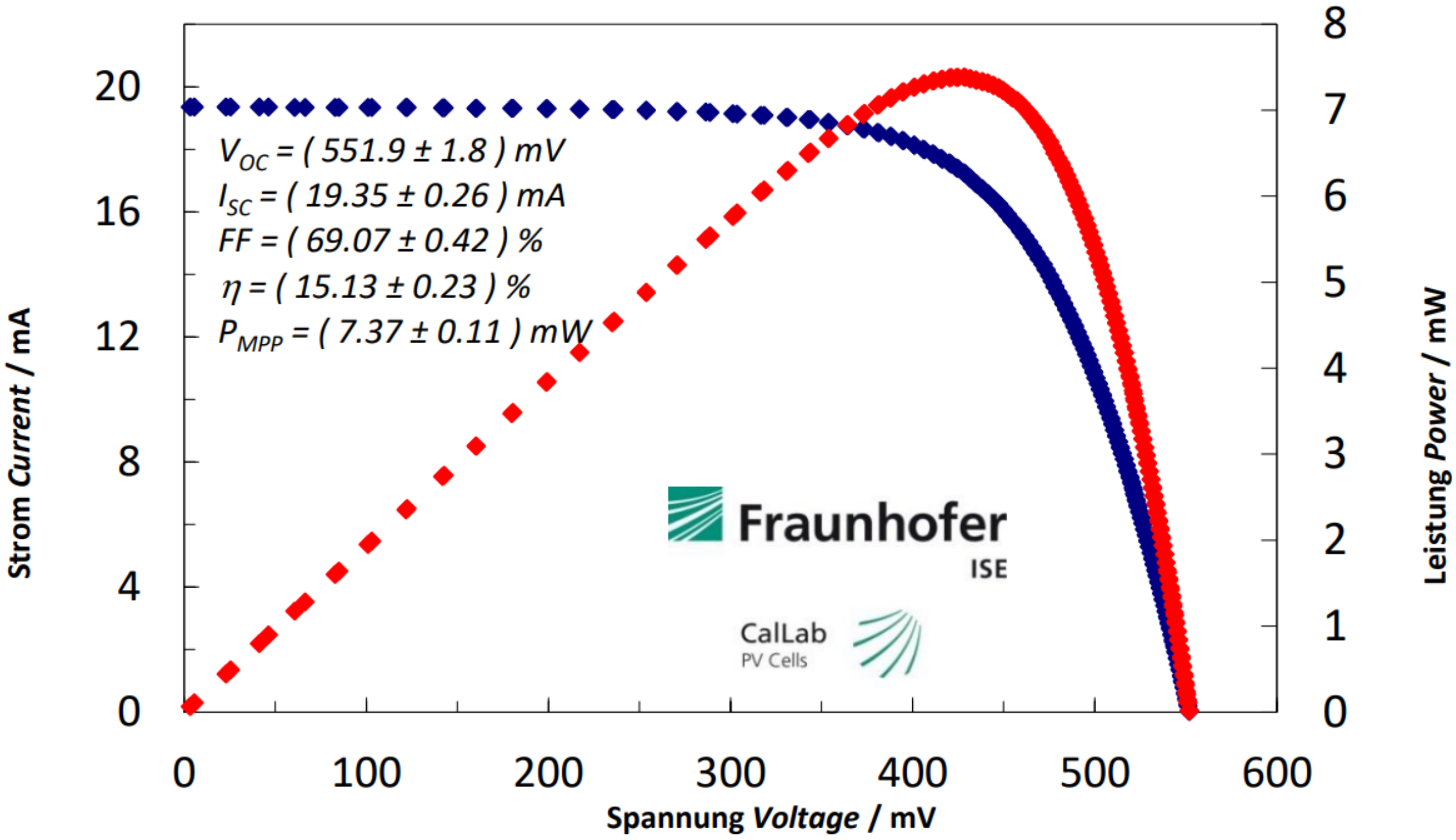

**Figure S22**. CISe HTL with Na barrier: bright and dark-field STEM image of a cross-section image near the backside (Mo at the bottom). All the layers of the HTL are visible. Below the columnar Mo layer, the 100 nm $SiO_x$ sodium-blocking layer is observed above the SLG substrate, as expected.

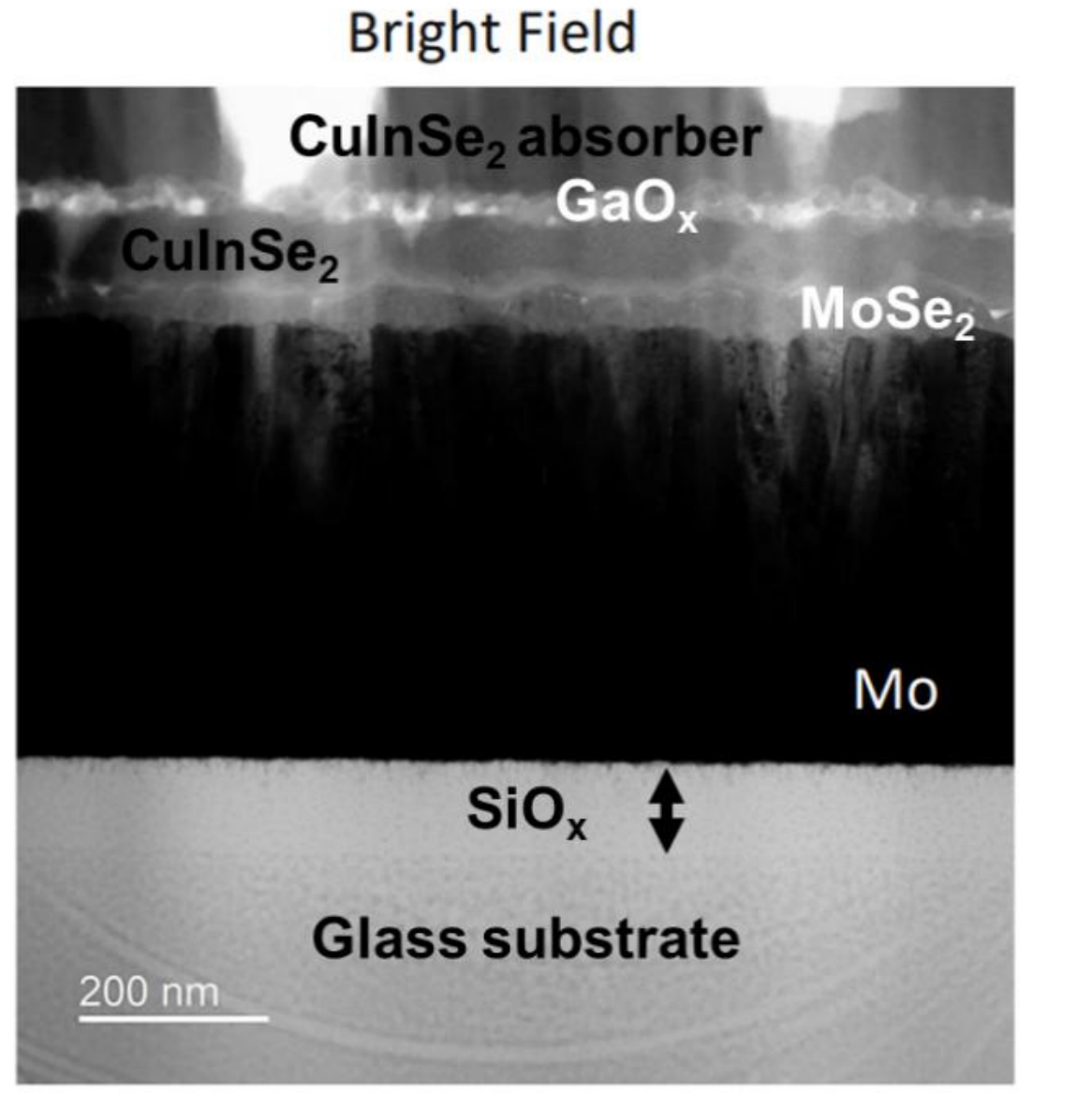

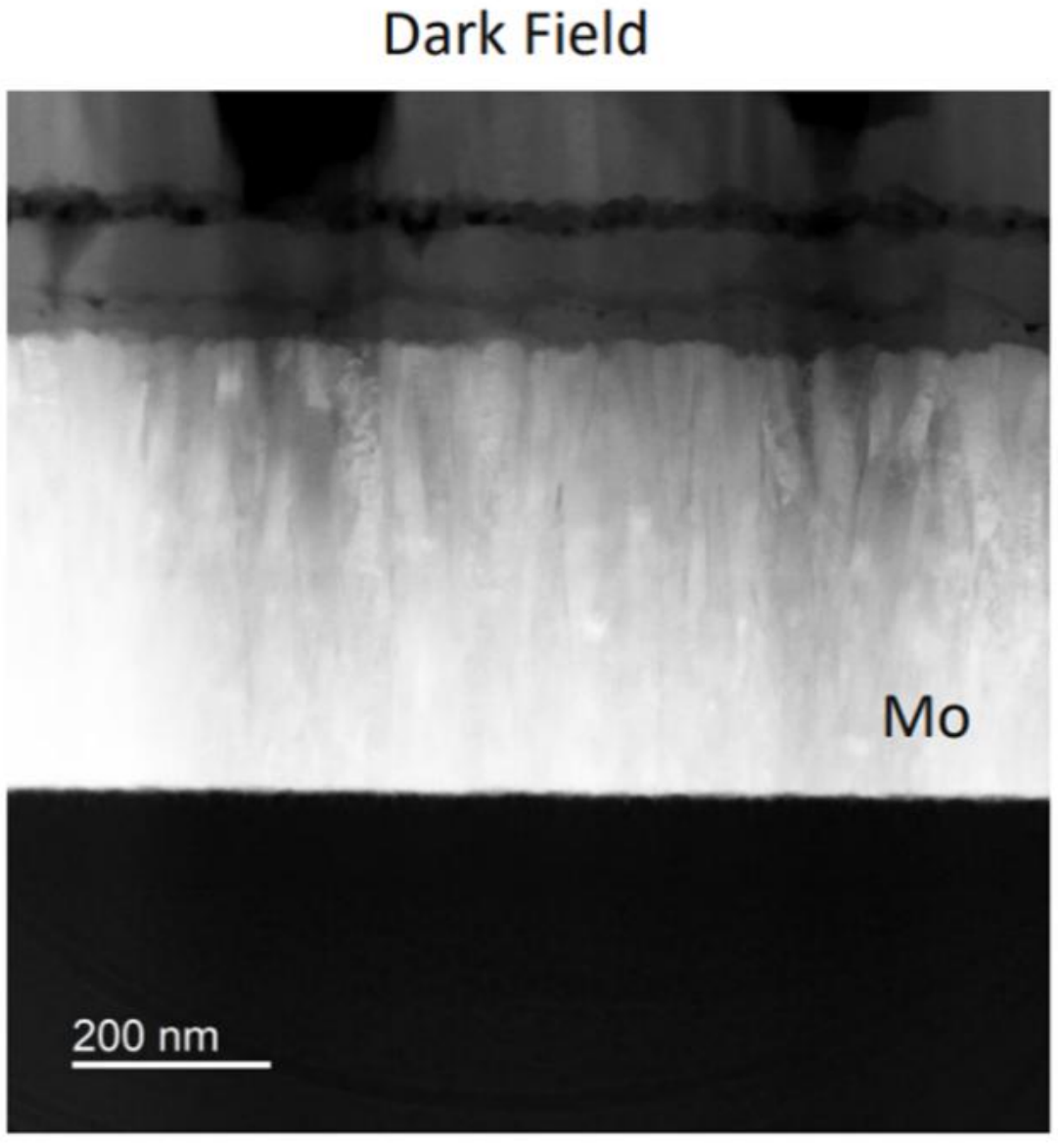

**$InO_x$ recipe development for 1.0 eV $CuInSe_2$ bottom cells.**

In this work, we have explored the possibility to optimize the hole transport layer (HTL) by modifying the parameters of the $InO_x$ deposition recipe. The $InO_x$ layer was initially sputtered from an $In_2O_3$ target onto approximately 120 nm of $CuGaSe_2$, which later led to In–Ga ion-exchange. The reference deposition was performed at 130 W and 1 mTorr Ar for 6 minutes, without additional substrate heating, oxygen supply or applied substrate bias voltage. The resulting $InO_x$ film was measured as approximately 30 nm via ellipsometry. Four groups were examined by varying only one parameter in each case, while keeping all other conditions constant as shown in **Table S1**. Note that all depositions were carried out for 6 minutes. Thickness optimization is reported elsewhere.[14]

**Table S1.** $InO_x$ Sputtering Recipe Group

| Group | Varying Parameter | Constant Parameters |
|---|---|---|
| a | Power (W): 110, 120, 130, 140. | 0 V, 1 mTorr, 0 sccm $O_2$ |
| b | Pressure (mTorr): 1, 2, 3, 5. | 0 V, 130 W, 0 sccm $O_2$ |
| c | Applied Substrate Bias (V): 0, 10, 15, 25. | 1 mTorr, 130 W, 0 sccm $O_2$ |
| d | Oxygen Supply (sccm): 0, 0.5, 1, 10. | 0 V, 1 mTorr, 130 W |

Based on ellipsometry measurements of the $InO_x$ films deposited on Si reference samples, all films exhibited a thickness of 30 ± 5 nm, except for the samples with 1 and 10 sccm $O_2$ in Group 4, which had thicknesses of 37.4 nm and 13.5 nm, respectively.

**Figure S23. a-d** present the J-V comparisons of the final devices in the corresponding groups in Table S1. Note that the yellow lines in each group represent the reference solar cell. Variations in sputtering power, pressure, applied bias, and oxygen supply were observed to only slightly affect the solar cell performances without any clear trend. For instance, 120 W and 140 W (higher and lower than the reference) as shown in a) provide slightly better $V_{oc}$ than the reference cell (130 W); increased pressure 5 mTorr as illustrated in b) offers a somewhat better FF than the reference cell (1 mTorr), but 3mTorr leads to a considerably lower *FF*; 15 V applied bias provides a better $V_{OC}$ than the reference cell (0 V) as demonstrated in c); oxygen flow during sputtering significantly decreases *FF* as d) shows. Nevertheless, no single sputtering condition leads to a consistent improvement in overall device performance. And no trends were observed.

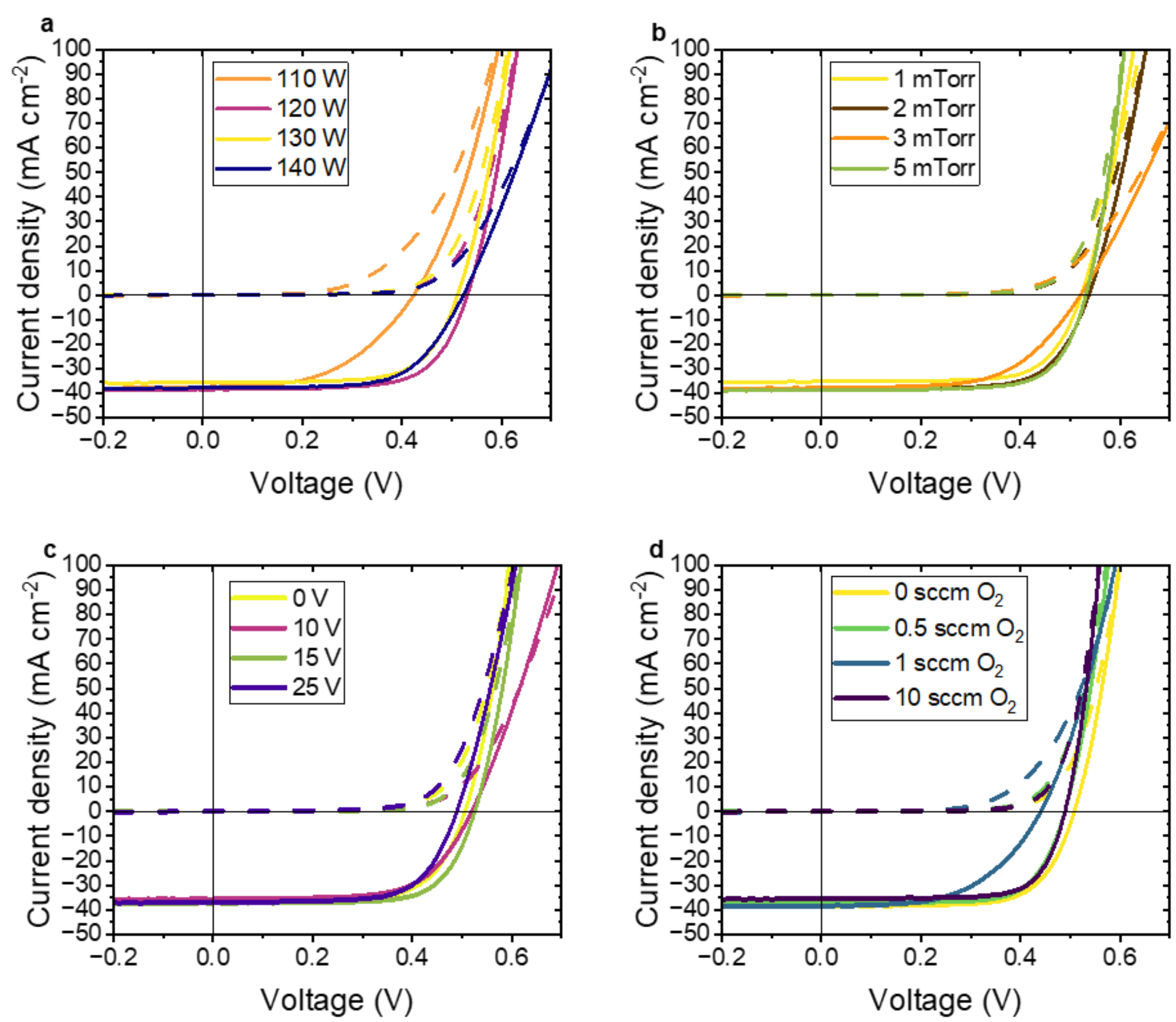

**Figure S24**. SEM images of $CuGaSe_2/InO_x$ stack top view.

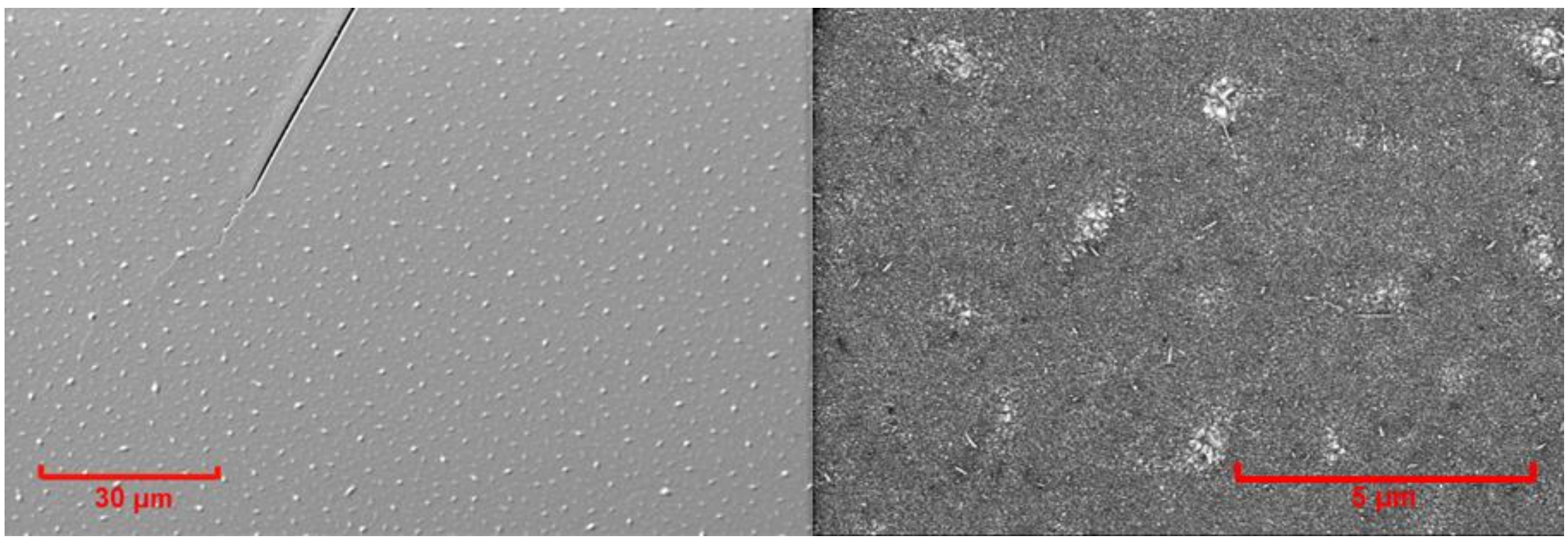

Additionally, we observed cracks and lumps on all the thin $CuGaSe_2$ layers, and they remained visible after the $InO_x$ deposition under SEM as shown in Figure S24. Two images at different scales. One example of reference sample (with $In_2O_3$ sputtering condition of 130 W, 1 mTorr, no oxygen flow or applied bias) is demonstrated in Figure S24.

Starting from the standard process, which includes Mo cleaning (a 500 °C annealing step of Mo substrate) and a 3-minute Se pre-supply step, several process modifications were investigated to assess the formation of the cracks and lumps. These modifications included the removal of the Mo cleaning step, the removal of the Se pre-supply step, and adding a post-anneal at 500 °C after $CuGaSe_2$ growth. The corresponding effects on the final device performance were also evaluated as illustrated in Figure S25. These modifications did not eliminate the cracks and lumps, and we find Mo cleaning prior to $CuGaSe_2$ growth is necessary for good performance of finished 1.0 eV $CuInSe_2$ solar cells, as illustrated in **Figure S25a**. Additionally, the comparison in **Figure S25b** shows that adding post-anneal has almost no impact on the finished solar cells.

**Figure S25**. *J-V* characteristics of 1.0 eV $CuInSe_2$ solar cells with different growth conditions of $CuGaSe_2$ precursors.

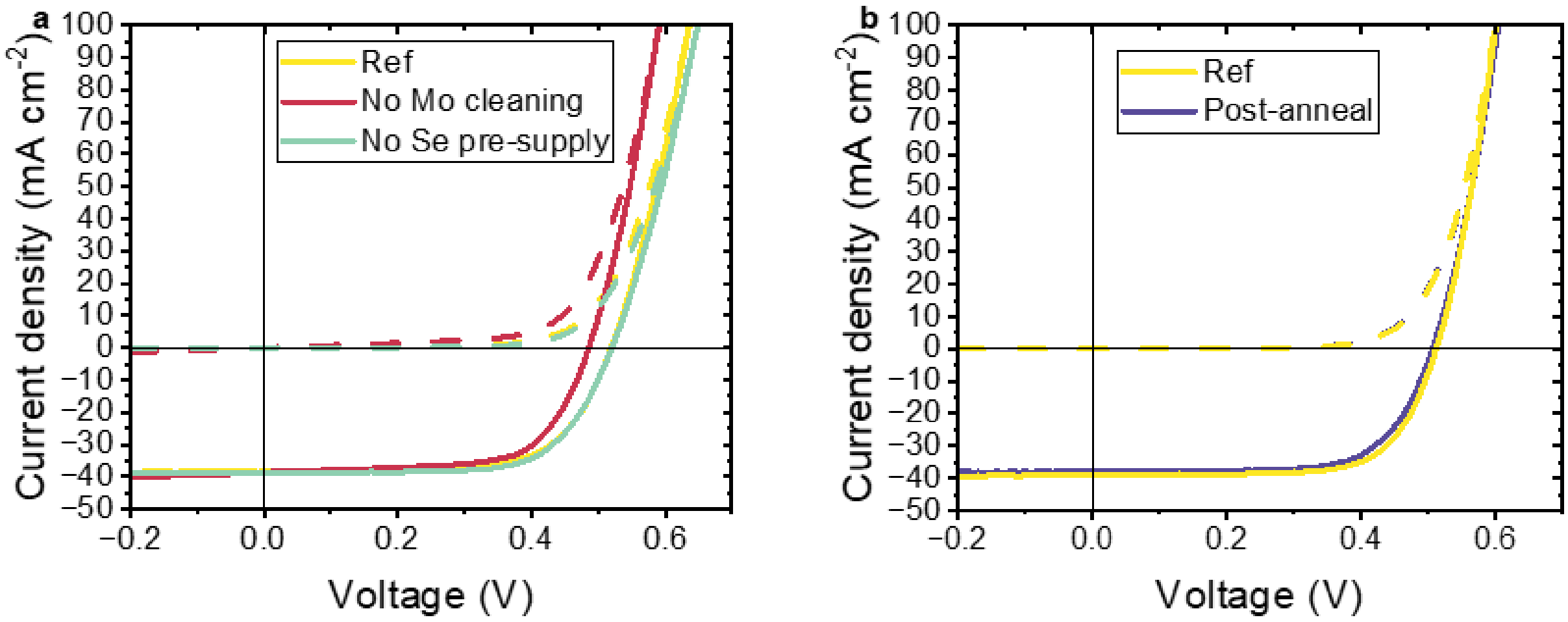

To summarise, no clear trend was identified among the sputtering conditions that would lead to improved hole transport layer properties or enhanced solar cell efficiency. Although cracks and lumps were observed on the $CuGaSe_2$ layer, they did not appear to impact the HTL or overall device performance. Most importantly, we clearly see the passivating effect of the HTL, even though the $CuGaSe_2$ layer has cracks and lumps.